# A review about wüstite $Fe_{1-z}O$, pseudo-phases and defect clustering

**Jean-Raymond Gavarri [1*] and Claude Carel [2]**

[1] *Institut Matériaux Microélectronique et Nanosciences de Provence, IM2NP, UMR CNRS 7334, Université de Toulon, BP 20132, 83957 La Garde Cedex, France*
* *to whom correspondence should be addressed* - e-mail = gavarri.jr@univ-tln.fr (or gavarri.jr@gmail.com)

[2] *Professeur honoraire de l'Université de Rennes1, Rennes, France*
   *27 cours Raphaël Binet, 35000 Rennes, France* - e-mail = c-carel@orange.fr

*February 2018: Second version*
(Preceding version (Dec. 2017): J.-R. Gavarri, C. Carel, same title, 'Document de travail' 2017 <hal-01659994> )

*Abstract*

Thermodynamic properties and structural aspects of the nonstoichiometric wüstite $Fe_{1-z}O$, and its modifications - the so-called pseudo-phases - as functions of composition $z$ and of equilibrium temperature are reviewed from 1960 to present. The complexity of the equilibrium phase diagram is described in details. The first order transition W ⇆ W' is specified on the iron/wüstite boundary near 1185 K. Transitions correlated to the modifications Wi at $T(W) > 1185^{++}$ K and W'j at $T(W') < 1185^{+}$ K (i and j =1,2,3) are re-examined. Structural determinations based on the characterization of point defects stabilization and of their clustering are reviewed. Additionally, the pseudo-phases are examined based on the transformation of defect clusters or of their mode of distribution (*i.e.,* percolation or superstructure) with the inclusion of changes in electronic charge carriers.

*Résumé*

Une mise au point bibliographique des propriétés thermodynamiques et structurales de la wüstite non-stœchiométrique $Fe_{1-z}O$ et de ses variétés -ou *pseudo-phases*- est faite de 1960 à ce jour. L'attention est d'abord portée sur la complexité du diagramme de phases à l'équilibre. La transition du premier ordre W ⇆ W' est précisée sur la frontière fer/wüstite au voisinage de 1185 K. Les transitions dues à l'existence des pseudo-phases Wi à $T(W) > 1185^{++}$ K et W'j à $T(W') < 1185^{+}$ K (i et j =1,2,3) sont reconsidérées, de même que les déterminations structurales relatives aux défauts ponctuels et à leurs amas. Des équilibres sont envisagés, qui tentent de caractériser la stabilité des pseudo-phases. Celles-ci peuvent être interprétées en termes de transformation ou de mode de distribution des amas de défauts (*i.e.,* percolation, surstructure) en incluant des changements dans les porteurs de charges électroniques.

## I – Introduction

The nonstoichiometric iron monoxide $Fe_{1-z}O$ is the main constituent of traditional blast furnace slags. Historically, this oxide was the subject of numerous studies in the general context of iron metallurgy improvement, and also for its role in the fields of recycling of industrial waste, catalysis and nanoparticles.

The oxide $Fe_{1-z}O$, also written equivalently as $Fe_yO$ or $FeO_x$, is named after the German minera-logical appellation "wüstit"[1]. It is considered of particular interest by the geophysicists because it is a constituent of the lower mantel of earth (*See particularly* Hazen and Jeanloz's review (R-84Haz)).

This oxide is in a steady state, between Chaudron's point C near 600 °C (21Cha) and its melting temperature around 1400 °C (45Dar), under an equilibrium oxygen pressure in the range $10^{-25}$ to $10^{-6}$ bar. This oxide exists with a large iron deficiency, which is expressed by the nonstoichiometry departure $z \in [0.04 - 0.18]$ around 1400 °C and $z \cong 0.065$ near 600 °C.

A number of reviews or equivalent papers on the nonstoichiometric wüstite was published previously: Goodenough (R-71), Per Kofstad (R-72), Vallet (R-72), Burgmann (R-75), Gokcen (R-75), Men' *et al.* (R-77), Spencer and Kubaschewski (R-78), Bauer and Pianelli (R-80), R-Toft Sørensen (R-81), Hazen and Jeanloz (R-84), Gleitzer and Goodenough (R-85), Gokcen (R-86), Lykasov *et al.* (R-87), Mrowec and Podgoreka (R-87), Tomlinsom *et al.* (R-90), Long and Grandjean (R-91), Sundman (R-91), Wriedt (R-91), Gleitzer (R-97), Smyth (R-2000), Desré and Hodaj (R-10), Worral and Coley (R-13).

Multiple studies have addressed the nature of the point defects (iron vacancies and interstitials) in wüstite, and the concept of defect clustering was systematically developed with a large variety of structural models. The NaCl face centered cubic structure of wüstite is characterized by a high rate of iron vacancies, and the existence of a certain proportion of iron ions occupying interstitial sites. Crystallographic determinations showed that these defects are not disordered in the lattice. Diverse models of clusters of iron vacancies and interstitials have been proposed. Given all the data available to date, a more accurate chemical formula (1a) can be expressed using classical ionic notations as follows

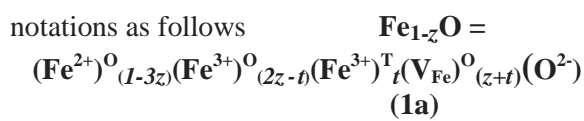

(1a)

---
[1] There is no equivalent available in the English language for the German phoneme 'ü'.

In this expression, the ionic species $(Fe^{2+})^O$, $(Fe^{3+})^O$, $(Fe^{3+})^T$ and $(V_{Fe})^O$ are characterized by fractions $(1-3z)$, $(2z-t)$, $t$ and $(z+t)$ respectively for one atom of ionized oxygen $O^{2-}$. Iron vacancies on octahedral sites are assumed to be neutral and noted $V_{Fe}$ as a first approximation. The superscripts O and T $((-)^O$ and $(-)^T)$ designate octahedral and tetrahedral sites of the NaCl structure, respectively. The subscripts $z$ and $t$ are the fractions of iron vacancies and $Fe^{3+}$ occupying tetrahedral sites, respectively. Because it simplifies the structural description of defects, the oxygen site is assumed to be fully occupied (*i.e.*, there is no significant amount of oxygen vacancies). Adopting the international notation of Kröger and Vink, formula **(1a)** is written as follows:

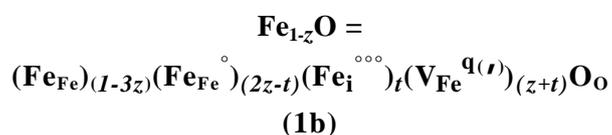

(1b)

In this expression, iron vacancies can be charged corresponding to q=2 or q=1, or neutral corresponding to q=0. Electric neutrality would result from electrons jumping in the conduction band, or interacting with holes in the valence band. It should be recalled that the band gap of wüstite was recently evaluated at about 1.0 eV at 25 °C, which means that a large population of electrons in the conduction band and/or holes in the valence band can be expected at high temperature, depending on the oxygen partial pressure (12Sch).

In this paper, we review historical observations and concepts for wüstite at equilibrium and in quenched samples. Then, we show how the complex non-stoichiometry of wüstite can be better described using mixtures of point defects and their associations in clusters.

## II - The complex behavior of wüstite

### II.1. Early work (1921-1961)

**1 -** In 1921, G. Chaudron's work in H. Le Chatelier's lab determined experimentally with a very high accuracy the external boundaries of the stable domain of ferrous oxide considered to be the daltonide 'FeO' (21Cha). Point C at the intersection of the two boundaries, where three solid phases (Fe, $Fe_3O_4$ and 'FeO') coexist at equilibrium at 570°C -temperature that has been revised in particular by (86Val)- is named Chaudron's point.

**TABLE I.1 - Notations and symbols presently used**

|  | Formal ionic notations | Kröger and Vink notations |
|---|---|---|
| $Fe^{2+}$ in octahedral site of the fcc NaCl-structure | $[Fe^{2+}]^O$ or $Fe^{2+}$ | $[Fe_{Fe}^x]$ or $Fe_{Fe}$ |
| $Fe^{3+}$ in octahedral site of the fcc NaCl-structure | $[Fe^{3+}]^O$ or $Fe^{3+}$ | $[Fe_{Fe}^\circ]$ or $Fe_{Fe}^\circ$ |
| $Fe^{3+}$ in tetrahedral site of the fcc NaCl-structure | $[Fe^{3+}]^T$ or $Fe_i^{3+}$ | $[Fe_i^{\circ\circ\circ}]$ or $Fe_i^{\circ\circ\circ}$ |
| Iron ($Fe^{2+}$) vacancy in octahedral site |  | $[V_{Fe}^{q(\prime)}]$ (q-times charged) |
| Iron ($Fe^{2+}$) vacancy in octahedral site |  | $[V_{Fe}'']$ (doubly charged) |
| Iron ($Fe^{2+}$) vacancy in octahedral site |  | $[V_{Fe}']$ (singly charged) |
| Iron ($Fe^{2+}$) vacancy in octahedral site |  | $[V_{Fe}]$ (neutral) |
| Oxygen on its site | $O^{2-}$ | $[O_O^x]$ or $O_O$ |
| Electrons and holes | $e^-$ ; $h^+$ | $e'$ ; $h^\circ$ |
|  | Other notations ||
| Partial pressure $p(O_2)$ / $\log_{10}(p(O_2))$  Partial pressure p(Fe gas) | $p'$ // $l'$  p ||
| $Fe_{1-z}$ O // $Fe_y$ O // $FeO_x$ // | $x = 1/(1-z) = 1/y$ ||
| 3D-superlattices | ➔ ($k_1.a \times k_2.a \times k_3.a$) with $k_1, k_2, k_3$, integers or not $\geq 1$  *(for a non cubic superlattice with 3 different cell parameters)*  ➔ (k.a) with $k_1 = k_2 = k_3$   *(for a cubic superlatttice)* ||
| References | (21Cha) = article, 1921, Chaudron  - Gleitzer (R-97) = Gleitzer, Review article (1997)  - (R-75Bur) = Review article (1975) Burgmann ||

**2 -** Jette and Foote demonstrated by means of measurements of the cell parameter and density after quenching that wüstite was an iron deficient solid solution: an iron vacancy being « a group FeO substituted by O ». They deduced that for the ionic compound, a missing $Fe^{2+}$ was necessarily compensated by the oxidation of two other $Fe^{2+}$ giving rise to electron holes $Fe^{3+}$ (33Jet).

**3 -** Darken and Gurry studied in details the phase diagram when determining the composition of wüstite chemically after slow quenching from T = 1100, 1200, 1300 and 1400°C. The boundaries with iron and magnetite were determined. The corresponding lines down to Chaudron's point were thus obtained from a far range extrapolation. Additionally, the authors determined the molar partial and integral thermodynamic properties of wüstite (45- 46Dar).

**4 -** Todd and Bonnickson, and Humphrey *et al.* (51Tod) (52Hum) each determined the values of entropy to be 59.4 and 57.5 J.K$^{-1}$. mol$^{-1}$, respectively, differing in 1.9 units for the same quenched wüstite Fe$_{0.947}$O. The former determinations considered a calculated configurational entropy of 1.72 J.K$^{-1}$ on the assumption of a random distribution of iron vacancies. They deducted that the lacunary iron lattice envisaged by Jette and Foote (33Jet) was ordered.

**5 -** Brynestad and Flood observed that some (electron holes) Fe$^{3+}$ are necessarily located in tetrahedral sites written as [Fe$^{III}$O$_4$] (58Bry).

**6 -** Using neutron diffraction analyses, Roth determined the proportion of Fe$^{3+}$ occupying the tetrahedral sites of the fcc structure and concluded to the formation of clusters (2:1) constituted of two iron vacancies and one tetrahedral site occupied by a cation Fe$^{3+}$, also named Frenkel cluster (60Rot). Tarte *et al.* also showed directly by I.R. spectroscopy the existence of these Fe$_\mathbf{i}^{\circ\circ\circ}$ ions (69Tar).

**7 -** Using the formalism of the mass action law initially proposed by Carl Wagner *et al.* (36-52Wag), Brynestad and Flood (58Bry), and subsequently Smyth (61- R-00Smy) established a simplified expression linking the composition $z$ to the oxygen partial pressure as $z \propto p'^{1/5}$ to describe the defect structure of both Mn$_\mathbf{1-z}$O and Fe$_\mathbf{1-z}$O, based on the experimental compositional data from Darken and Gurry (45Dar).

## II.2 Modifications or pseudo-phases apparent in the phase diagram

### 1 - Raccah -Vallet's team

During years 1962-65, three supposed « allotropic varieties » or « modifications » noted Wi (i=1,2,3) named later « subphases » and then « pseudo-phases » by other authors were discovered by Raccah *et al.* (62Rac**1,3**) (63- 65Val, 65Kle) from experimental measurements in Raccah's doctoral thesis (62Rac**2**). An initial layout of the corresponding phase diagram was proposed, based on coordinates (Θ°C, *x*) (64Val). Raccah's results were re-interpreted thermodynamically by Kléman (65Kle) in terms of a similar phase diagram.

In this review, the term pseudo-phase is used to design a structural and/or electrical configuration of wüstite, and the term boundary is used to designate the separation between two domains in which each major pseudo-phase is stabilized, but is not used to refer to the external limits of the whole domain.

During their thermogravimetric and thermodynamic studies, Vallet *et al.* (64- 65- 70- 75**1**- 75**2**- 79- 86- 89Val) established correlations between mass variation, *i.e.* departure from stoichiometry expressed by *x* in FeO$_x$ (*See* Table I.1: $y = 1/x$ in Fe$_y$O, $z=(x-1)/x$ in Fe$_\mathbf{1-z}$O), temperature T and oxygen partial pressure p′ under equilibrium (*See also* (R-72Val) p. 277 and 294-295). They described the phase diagram of wüstite FeO$_x$ above 911°C (*See* below for domain of W) through the following three equations connecting log p′, noted *l*′ (*See* Table I.1), with the composition *x* and temperature T, and expressing the three behaviors of pseudo-phases W1, W2 and W3 (*See* (89Val), Table 1 p. 210), successively:

$$l' = (46753.4T^{-1} - 7.3781)x + (-78825.3T^{-1}+16.0613) \quad \text{(II-2a-W1)}$$

$$l' = (-9568.9T^{-1} +31.1728)x + (-18413.3T^{-1} - 25.2569) \quad \text{(II-2b-W2)}$$

$$l' = (-33238.9T^{-1} + 48.3669)x + (6883.9T^{-1} - 43.5669) \quad \text{(II-2c-W3)}$$

To better relate the non-stoichiometry and equilibrium pressure p′ (*in* relationships II-2a, II-2b and II-2c) to the presence of defects (mainly cation vacancies and interstitials), we have expressed *l*′ as a function of *z* instead of *x* and we have reported *l*′ as a function of log *z* values. It should be recalled that these relationships derive from the general expression of equilibrium constants K(T) linking the composition *z* to the oxygen partial pressure p′. A basic setup can be proposed as follows:

$$\tfrac{1}{2} O_2 \leftrightarrows O_O + V_{Fe}^{q(\prime)} + q\, h^\circ$$

with $\quad [h°] = q.[V_{Fe}^{q(\prime)}] = q.z \quad$ (II-3a)

$$K(T) = \{[O_O].[V_{Fe}^{q(')}].[h°]^q\} / (p')^{1/2} \quad \text{(II-3b)}$$

$$K(T)^2 = C^2 \cdot z^s / p' \rightarrow$$

$$\log z = (1/s) \cdot l' + (1/s) \log(K/C)^2 \quad \text{(II-3c)}$$

where C depends on q.

In these equations, the activity $[O_O]$ of oxygen on its site fcc is considered equal to 1 because no significant amount of vacancies can be present in the oxygen lattice. The equilibrium constants depend on the values 2, 1, and 0 of q. The three hypothetical corresponding values of exponent s should be 6, 4, and 2 (more likely close to these values, taking into account the various approximations made). In the remainder of this review, exponent s is taken as characteristic of the specific equilibrium equation connecting the fractions of point defects between them (*See below* Section IV).

It should be noted that these empirical relationships are not strictly linear. They assume the existence of three domains, and allow a comparison with classical models of defect equilibria. In each domain, a unique model of defect equilibrium is supposed to be valid, which is not strictly exact because of the continuous evolution of interactions between defects, as composition and temperature vary in the phase diagram.

**Fixed temperature and variable composition**

To illustrate the different behaviors or transitions in Figure II-1 below, we detail the relation log *z* vs *l'* in the specific case of Vallet *et al.* data [89Val] as obtained at 1000 °C (T=1273 K) by transformation of the relations (II-2a, b, c) above and yielding:

$$\log z = 0.2036 \, l' + 1.7511 \rightarrow s = 4.9 \, (\pm 0.10)$$
$$\text{(II-4a-W1)}$$

$$\log z = 0.1369 \, l' + 0.8352 \rightarrow s = 6.7 \, (\pm 0.12)$$
$$\text{(II-4b-W2)}$$

$$\log z = 0.1124 \, l' + 0.5246 \rightarrow s = 8.9 \, (\pm 0.15)$$
$$\text{(II-4c-W3)}$$

From the three segments observable on the curve *l'* vs log *z* in Fig. II-1, we can derive the three values s(W1) = 4.9, s(W2) = 6.7, and s(W3) = 8.9. These values are similar to the ones found by Toft Sørensen (85Sor), Rekas and Mrowec (87Rek).

In reality, there is no reason for assuming linear relationships given the probable existence of mixtures of defects, and of their continuous evolution conditioned by the changes in charge of $V_{Fe}^{q(')}$ (q=0,1,2).

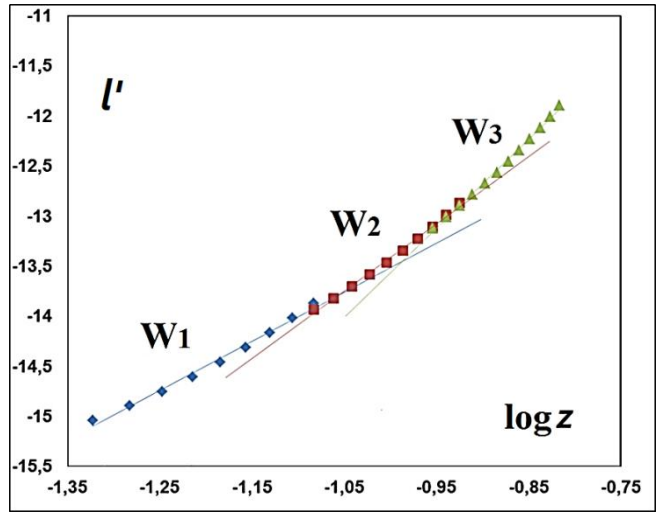

**Fig. II-1** - The logarithm of oxygen partial pressure p′ *vs* the logarithm of composition *z* derived from the data by (89 Val) at 1273 K. Three slopes can be determined giving rise to the three values of exponent s.

**Fixed composition and variable temperature**

For a fixed value *x* = 1.13 (*z* = 0.115), we report in Figure II-2 the experimental values of *l'* versus 1/T using the data of Vallet *et al.* (89Val).

Three quasi linear domains can be distinguished suggesting the existence of three Gibbs free energies of reaction in relation with the equilibrium constant K(T), with the classical relationship ΔG(reaction) = ΔH – TΔS = - RTlnK.

The linear correlations are:

$l' = -2.5994(10^4 \cdot T^{-1}) + 7.724$ (II-5a-W1)

$l' = -2.9226 (10^4 \cdot T^{-1}) + 9.9684$ (II-5b-W2)

$l' = -3.0676 (10^4 \cdot T^{-1}) + 11.088$ (II-5c-W3)

Using the simplified relationship without activity coefficients (61Smy):

$$K(T)^2 = C^2 \cdot z^s / p'$$ (II-3c)

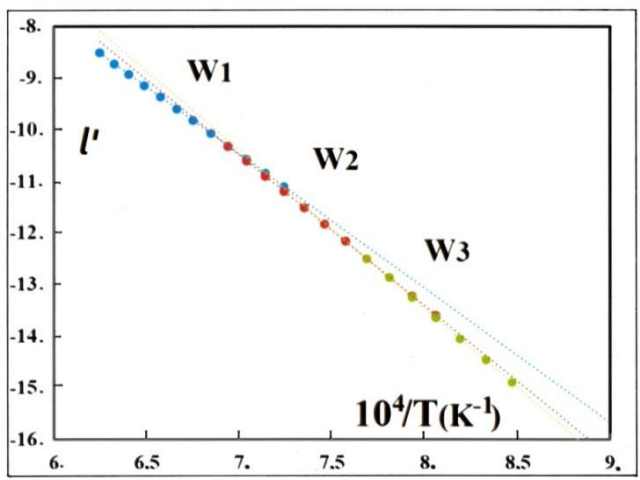

we can derive the following relations for the fixed value $z = 0.115$

⇨ $\log = (s/2) \cdot \log z - \tfrac{1}{2} l' + \log C =$
  $- (2/2.3026) \Delta G/RT$ (II-6c)
⇨ $dl' / d(1/T) = (2/2.3026) \Delta H/R$ (II-7c)

From this last relationship, we can determine three values of enthalpies linked to the changes in slopes and attributable to the so-called pseudo-phases

W1: $\Delta H_1/R = - 25526$ K  ➡  $\Delta H_1 = - 2.20$ eV
W2: $\Delta H_2/R = - 29322$ K  ➡  $\Delta H_2 = - 2.52$ eV
W3: $\Delta H_3/R = - 31008$ K  ➡  $\Delta H_3 = - 2.67$ eV

Fig. II-2 - Variation of *l'* *vs* 1/T for a fixed value $z = 0.115$: three quasi linear correlations corresponding to the initial determinations in the three domains W1 (high T), W2 (intermediate T) and W3 (low T) give rise to three activation energies $\Delta H_i$.

At the same time, a qualitative dilatometry study under CO2/CO indicated weak changes of slopes in curves $\Delta \ell(T)$ with intersections located on the boundary W2/W3 (64Car). In addition, several correlations between the cubic cell parameter obtained after quenching or *in situ* and the composition $x$ in $FeO_x$ led to a possible separation in three pseudo phases or subdomains (65- 76Car) of the equilibrium phase diagram.

In the equations above, we have neglected the activity coefficients, assuming that the ionic interactions remained constant as $z$ varies. Another point of view would be to consider that a significant evolution of the activity coefficients can be observed and that the variation is associated with molar fractions of point defects «i» at equilibrium (the activity $a_i$ being proportional to the molar fraction [i] : $a_i = \gamma_i \cdot [i]$). In the case of ionic solids, these activity coefficients $\gamma_i$ are generally used to express deviations from ideal models of isolated defects at equilibrium. Nowotny and Rekas (89Now) assumed the existence of only one hypothetical specific model, the negatively charged cluster $(4:1)^{5-}$. They derived a mean value in a simplified form for coefficient $\gamma_\pm$ using the Debye-Hückel model restricted to the case of limited development, and considered purely ionic interactions only. Such theoretical variations of activity coefficients as a function of $z$ corresponds to a continuous description across the wüstite phase diagram. However, such a description is not suitable due to the existence of the pseudo-phases.

**3 - Abnormal behavior near 1184/1185 K and suggested diagram**

A splitting of the phase diagram was first proposed by (70Val) and then was described more precisely (75- Val**1** and Val**2**). The splitting concerned the two subdomains for phases W' and W, for each part of the first order transformation α-Fe ⇆ γ-Fe, each subdomain being constituted of three pseudo-phases W'j and Wi, respectively. Because the properties of W' and W are distinct, the boundaries with α- or γ-iron of the stability domains are also distinct. That is the reason for the existence of points B and A at the upper and lower end respectively of the corresponding curves in the phase diagram (*See* (86Val), Tables IV & V p. 723 and 724). The limit compositions on these boundaries $y_{o(1)} = 0.9427$ or $z_{o(1)} = 0.0573$, and $y'_{o(1')} = 0.9528$ or $z_{o(1')} = 0.0472$ of W1 and W'1 in equilibrium with γFe-O$_\varepsilon$ and αFe-O$_{\varepsilon'}$ were firstly proposed at points A and B, respectively, both assumed at 1184 K, which cannot strictly occur as discussed further below.

In an extended careful review, **Wriedt (R-91Wri)** emphasized « *the extraordinary aspect of the continuous isothermal nature* at 1184 K *of the W'1/W1, W'1/W2, W'2/W2, W'3/W2 and W'3/W3 sub-boundaries, and their relationship, if any, to the α-γ Fe transformation* ». This pending question had no answer until the 1980s.

The basic equations governing the various equilibria between gases and solid phases (*See* equilibrium diagram in Fig. II-3*)* are listed below. At the two boundaries involving α–Fe$_{cr}$ or γ–Fe$_{cr}$ (subscript 'cr' for 'solid' *i.e.*, 'crystallized'), equilibria can be expressed:

1) on the boundary αFe/W'1 from point B to point C:

$(1-z_{o(1')})$ α–Fe$_{cr}$ + ½ O$_2$ ⇄ Fe$_{1-z_{o(1')}}$O ➔

$$K_\alpha(B) = \exp(\Delta G^o_\alpha / RT)$$

α–Fe$_{cr}$ ⇄ Fe$_{gas}$ ➔

$$K_{(\alpha,gas)} = p_\alpha(Fe_{gas})$$

$(1-z_{o(1')})$ Fe$_{gas}$ + ½ O$_2$ ⇄ Fe$_{1-z_{o(1')}}$O ➔

$K_{(\alpha, gaz)} = \exp -(\Delta G^o 1/RT) = 1 / (p_1(Fe_{gas}) \cdot p'^{½}_\alpha)$

2) similarly on boundary Feγ/W1 from point A to melting:

$(1-z_{o(1)})$ γ–Fe$_{cr}$ + ½ O$_2$ ⇄ Fe$_{1-z_{o(1')}}$O ➔

$$K_\gamma(A) = \exp(\Delta G^o_\gamma / RT)$$

γ–Fe$_{cr}$ ⇄ Fe$_{gas}$ ➔

$$K_{(\gamma,gas)} = p_\gamma(Fe_{gas})$$

$(1-z_{o(1)})$ Fe$_{gas}$ + ½ O$_2$ ⇄ Fe$_{1-z_{o(1)}}$O   (2)

$K_{(\gamma, gaz)} = \exp -(\Delta G^o 2/RT) = 1 / (p_2(Fe_{gas}) \cdot p'^{½}_\gamma)$

At the arbitrarily unique temperature of 1184 K, it is thus necessary to apply two different partial pressures $p_1$ and $p_2$ of Fe(gas), at equilibrium with the solid at points A and B. Obviously, at the unique α⇄γFe equilibrium temperature, it exists only one value of the pressure p(Fe$_{gas}$) = 3.005x $10^{-11}$ atm. Thus, the points A and B cannot be at the same temperature.

NOTA: If T(A) and T(B) differ from a typical value of 4 °C (*See* reference (86Sjö) below), for example T(A) = 1191 K and T(B) = 1187 K, the partial pressures of iron are necessarily p(A)=p$_1$=3.692x10$^{-11}$ atm and p(B) = p$_2$=3.220x 10$^{-11}$ atm ((82Pan) p. 153 and (86Des) p. 967-83). Thus, the differences between the two iron partial pressures is on the order of 0.4x10$^{-11}$ atm. The corresponding partial pressures of oxygen are p'(A) = 4.829x10$^{-17}$ atm and p'(B) = 4.229x10$^{-17}$ atm, *i.e.* a difference of 0.6x10$^{-17}$ atm. The compositions $z_o(A)$ = 0.0571 and $z_o(B)$ = 0.0468 correspond to a gap of ~0.01 in composition.

Thus, in the case of the points A and B differing in temperatures T (A) and T (B), then the compositions $z_o(A)$ and $z_o(B)$, the partial pressures p'(A) and p(A), p'(B) and p (B) are necessarily different. Finally, a small difference in temperatures, probably a few degrees, should be determined experimentally in order to adjust this set of parameters, which determines the phase diagram (*See* Fig.II-3) close to stable temperature of the α-γFe transition.

In this situation, the bottom row in Table IV (86Val) should be changed to a value of T(A) higher than 1184/1185 K, and the header line of Table V (86Val) to a value of T(B) also higher than 1185 K but lower than T(A) (*See* (86Val) p. 723 and 724).

**Phenomena close to the isotherm at 1185 K**

Experimental confirmations of this hypothetical separation exist in the literature. Several of them are indicated below.

**Löhberg and Stannek** separated the pre-exponential term a and the energy term b = ΔG/RT of an expression functionally similar to a Boltzmann distribution pH2O/pH2 = a.exp(b.*z*) expressed as a function of T$^{-1}$ in two segments (*See* (75Löh) Fig. 8 and 9) intersecting in the vicinity of the α⇄γFe transition temperature.

Similarly, **Mrowec** *et al.* separated the determination of the self-diffusion coefficient of iron in wüstite as a function of $T^{-1}$ from their own results and several sets from the literature (*See* (81Jan) Fig. 7 p. 99). The intersection occurs for $T^{-1} = 8.28 \cdot 10^{-4}$ $K^{-1}$, *i.e.,* at 935°C, of the segments $D_{Fe}$ *vs* $T^{-1}$ attributed to W and W'.

**Jacobsson and Rosén** (81Jac), **Guillermet and Per Gustafson** (85Gui), **Sjödén** *et al.* (86Sjö), **Grønvold** *et al.* (93Gro) published new experimental thermodynamic data of highly improved accuracy about the molar thermodynamic properties of iron and wüstite.

**Sundman** in an assessment of the entire Fe-O system (R-91Sun) formulated a model for the 'pure iron corner' (solid solutions $\gamma Fe-O_\varepsilon$ and $\alpha Fe-O_{\varepsilon'}$). The assessment of **Spencer and Kubaschewski** (R-78Spe) is also relevant in the same context.

A departure from a monotonous variation of the electromotive force (e.m.f.) of galvanic cells, near the $\alpha \leftrightarrows \gamma Fe$ transition, was evidenced from (81Jac) measurements at the iron/wüstite electrode, in several galvanic cells, by **Sjödén** *et al.* (86Sjö). These latter authors published numerous determinations of the e.m.f. between 866.7 and 1339.3 K, measured in a galvanic cell equipped with a negative electrode constituted of the redox couple $Fe/Fe_{y_O}O$ (86Sjö). An anharmonic transitional zone between two separate continuous variations, not in continuation of each other, can be identified in the temperature range [1186-1194 K].

**Consequences relevant to the Fe-O phase diagram.**

The $\alpha \leftrightarrows \gamma Fe$ first order transition in the 'pure iron corner' of phase diagram Fe-O is not documented sufficiently. The left side of Figure II-3 reports literature data concerning the domain of this transition (67Swi) (R-78Spe) (R-91Sun).

The temperature of the strictly pure $\alpha$-Fe $\leftrightarrows$ $\gamma$-Fe transition is tabulated at 1184 ± 3 K in the data base NIST-Janaf ((98Cha) and 2011). At present, it is generally taken to be 1185 K (point ø in Fig. II-3).

The transition enthalpy is known with a 20% dispersion from assessed and experimental data, (*See* (85Gui): Table 3 p. 604). Very few experimental studies are available concerning the peritectoid invariant ($\alpha Fe-O_{\varepsilon'}$, $Fe_{1-z}O$, $\gamma Fe-O_\varepsilon$) close to the temperature of the pure $\alpha$-Fe $\leftrightarrows$ $\gamma$-Fe transition (67Swi). Sundman (R-91Sun) took up the molar fractions ($\cong \varepsilon$ at point P, $\varepsilon'$ at point b in Fig. II-3) of dissolved oxygen $N_O = x/(1+x) = 16 \times 10^{-6}$ and $6.8 \times 10^{-6}$ in $\alpha$- and $\gamma$-Fe respectively at 1185 K or 912 °C (Fig. II-3).

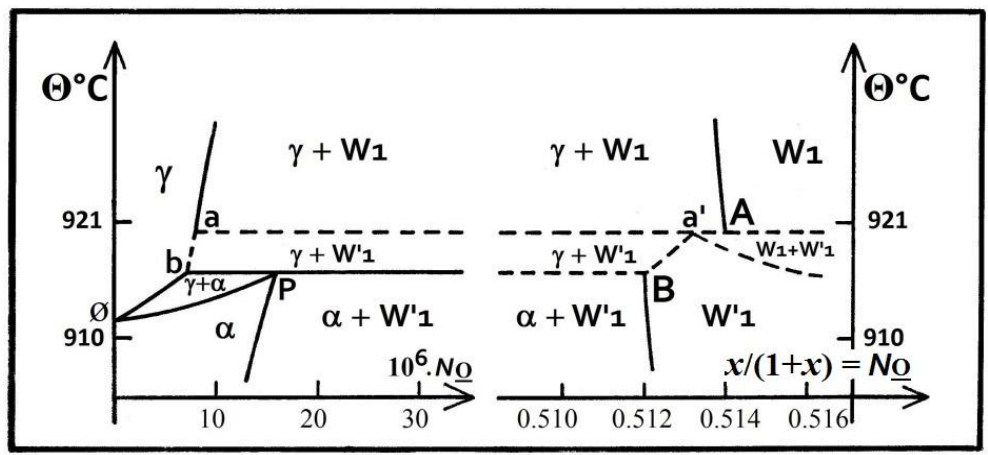

**Fig. II-3** - Equilibrium diagram between W1 and W'1 and 'pure iron', with both limit points A and B of $\gamma$-Fe/W1 and $\alpha$-Fe/W'1 boundaries. Compositions $N_O = x/(1+x)$ are arbitrarily taken at 1184 K in ((86Val) Tables IV and V). Dashed lines correspond with hypothetical limits.

The separation of the boundary iron/wüstite into parts $\gamma Fe-O_\varepsilon$/W1 from point A and $\alpha Fe-O_{\varepsilon'}$/W'1 from point B arises distinctly in the temperature range characterized in (87Sjö), following the phase diagram in the right part of Fig. II-3. Two types of two-phase domains (γ + W'1) and (W1 + W'1) have been added as well as an invariant point a' (W'1, γFe-O_ε, W1). As of now, the location of points A and B remains to be determined precisely. New experimental data is necessary to confirm the preceding suggestions.

For the subdomain of W' below 912 °C, the information equivalent to that of W above 912 °C is more difficult to access ((*See* (65Val) Figs. 6 and 7 p. 12 and 13). An empirical relationship associated with the three W'j requires a curvature term. The analytical expression is taken as follows:

$$l' = (a'_j T^{-1} + b'_j) x^2 + (c'_j T^{-1} + d'_j) x + (e'_j T^{-1} + f'_j)$$
(II-8)

The coefficients ($a'_j$ - $f'_j$) are available in ((89Val) Table 1 p. 210).

The sets of numerical relations detailed above for $l'$ as functions of $x$ and $T^{-1}$ allow the determination of the transition lines or boundaries between domains of the pseudo phases W1, W2 and W3 from γ-iron to magnetite between 912 and around 1400 °C, and W'1, W'2 and W'3 from α-iron to magnetite between 912 and 592 °C at point C, as shown in Figure II-4.

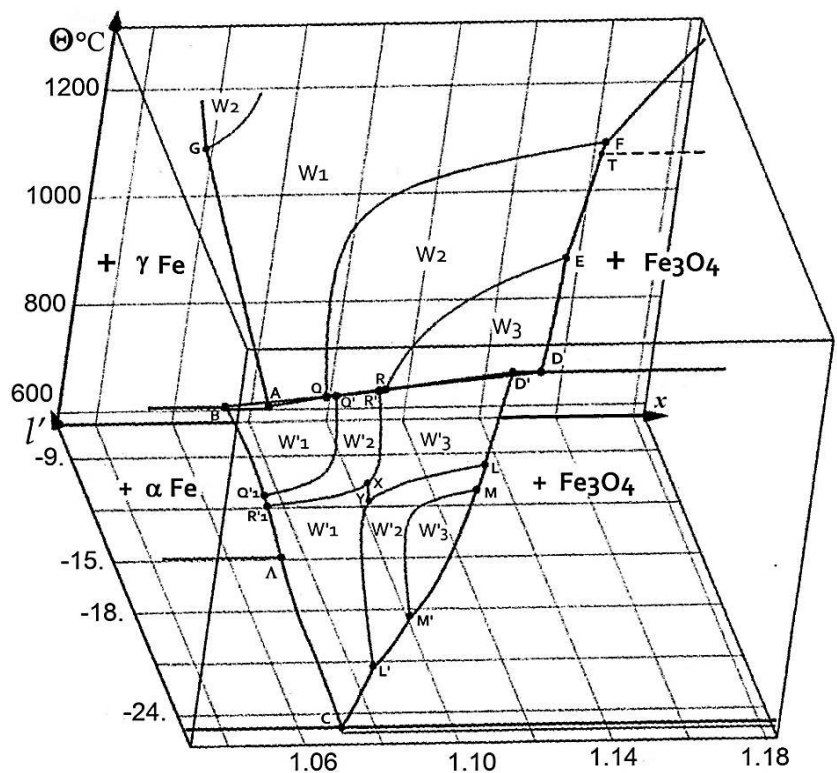

**Fig. II-4** – Phase diagram Θ°C(*l'*, *x*) of solid wüstite FeO_x under equilibrium following Raccah-Vallet-Carel (64-89Val): upper temperature 1250 °C, lower one 592 °C at point C; coordinate *l'* for $\log_{10}p(O_2)$. Points noted by a capital letter are tabulated in (79Val), except the point Λ in (86Val) and T in (81Car) (87Val) (12LiL).

## 4 - First evidences of phase modifications or pseudo-phases in the literature

**Wagner jr's team** In the course of thermodynamics - defect structure approaches by Carl Wagner (36- 52Wag), Wagner jr *et al.* identified some « *unexpected changes in the slopes of log σ vs. log $p_{O2}$* » of the electrical conductivity at equilibrium (*See* (66Gei) Table IV p.955; *also* Annex 1 a-), at controlled T and p'. They stated that there is « *a correlation between the estimated composition of the break of* [their] *curves and the compositions corresponding to Raccah et al. transitions* ». Using thermogravimetric data (68Swa) the transitions Wi/Wi+1 were identified at equilibrium outside of Vallet's group, for the first time. The further studies of Wagner's group about wüstite were concerned with transport phenomena in the bulk and on the surface in relation with the p to n transition.

**Fender's team**

From numerous accurate electromotive force measurements in galvanic cells, Fender and Riley (69Fen) brought outside of Vallet and Associates' group the second experimental confirmation of the likely existence of three pseudo-phases, and the location of the boundaries of the subdomains in the range 700-1350 °C. Their plot of the equilibrium diagram is in fairly good agreement with that of Vallet et al. (see Annex 1, Fig. A1-1).

Fender and Riley (69Fen), then later Vallet and Carel (86Val) assessed the partial molar properties of an oxygen solution ($\Delta \bar{H}_O \equiv \bar{H}_O - \frac{1}{2} H°_{O2}$, $\Delta \bar{S}_O \equiv \bar{S}_O - \frac{1}{2} S°_{O2}$) as functions of $y \in$ [0.945 - 0.853] for the three pseudo phases Wi named also wI, wII and wIII (*See* (69Fen) Table I p. 796, and in Fig. II-5 below). The following relations were formulated in (86Val):

$\Delta \bar{H}_O = 19.14475 (a_i.x + c_i)$,

$\Delta \bar{S}_O = -19.1447 (b_i.x + d_i)$         (II-9)

for the three Wi (i=1,2,3) at T >1184 K;

$(\Delta \bar{H}_O)' = 19.14475 (a'_j.x^2 + c'_j.x + e'_j)$,
$(\Delta \bar{S}_O)' = -19.14475 (b'_j.x^2 + d'_j.x + f'_j)$   (II-10)

for the three W'j (j=1,2,3) at T< 1184 K.

The coefficients $a_i - d_i$ given in § II.2. above 1184 K are those in the initial relation

$l' = (a_i \, T^{-1} + b_i).x + c_i \, T^{-1} + d_i$

The coefficients ($a'_j - f'_j$) below 1184 K are given in (89Val, *See* Table 1 p. 210).

Relations (II-5) for the Wi, and (II-6) for the W'j are independent of T, in accordance with the property of regularity of the solid solution wüstite identified in (62Rac**1**).

## 3 – Partial molar enthalpies and entropies. Convergence point Ω

$\Delta \bar{H}_O$ is represented in Figure II-9 from numerous available data sets directly evaluated or analytically derived such as in (69Fen), (70Mar) or (63-64-86Val).

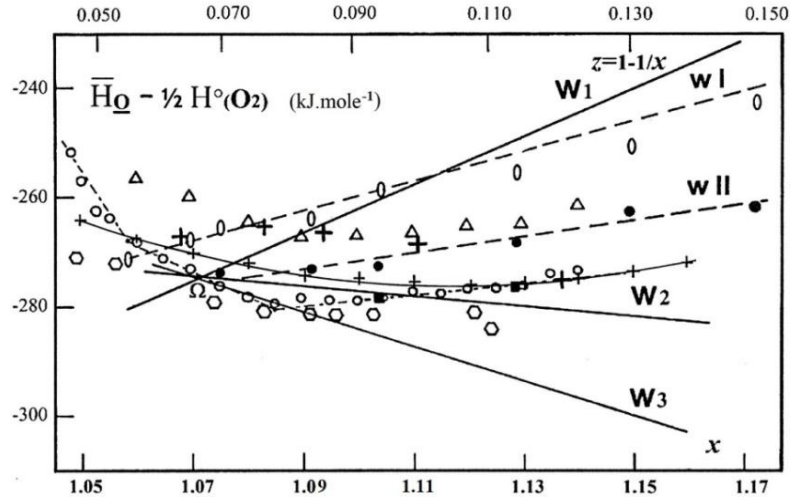

**Fig. II-5 -** $\Delta \bar{H}_O$ **from some authors.** ⎯ **: Derived by Vallet & Carel (86Val) for the Wi (i=1-3)** *i.e.* **at T ∈[1193-1523 K] from thermogravimetry + CO2/CO;** ◐--◐**: w I,** ●__●**: wII, and** ■**: wIII (two points only) by Fender and Riley (69Fen) for the three wI (I=I-III) from emf at T ∈[1023-1623 K]; dash-dot line ○-·-·-○: Gerdanian *et al.* (70Mar) from calorimetry at 1348 K under vacuum (+ δn_{O2});** ✚**: Asao *et al.* (70Asa) from emf at T∈[873-1273 K]; △: Takayama and Kimizuka (80Tak) from thermogravimetry + CO2/H2 at T ∈[1373-1673 K]; ○: Barbero *et al.* (81Bar) from emf at T ∈[823-1323 K]; ✚: Lykasov *et al.* (82Lyk) (R-87Lyk) emf data at T ∈[873-1373 K].**

A convergence in the vicinity of the common intersection point Ω at $x$ = 1.0713 (±0.001) of the straight lines $\Delta \bar{H}_O(x)$ from relations (II-5) is noticeable globally. Contrary to this contrived amalgamation process, the numerical adjustments presented next are related principally to the initial results from Raccah (62Rac2) (65Val), which form a well populated and coherent data set at equilibrium, and for which it exists now a systematic interpretation by Vallet's group (79- 86- 89Val), and indubitable confirmations (66Gei) (69Fen) (10Wor). The graphical representation of the relations (II-5) and (II-6) for the isotherms at 1000 and 820 °C, respectively, is given in Fig. II-6.

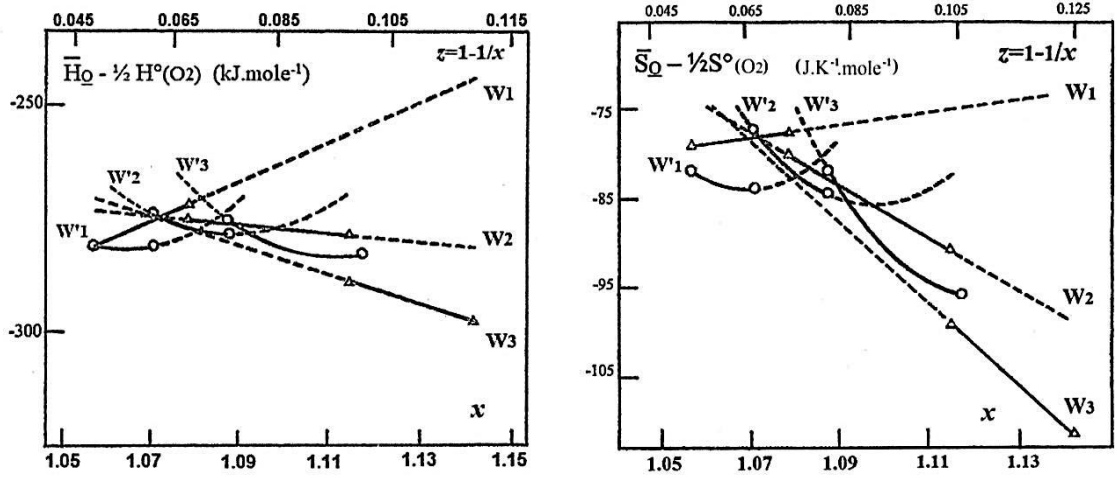

**Fig. II-6- Partial molar enthalpy ($\bar{H}_O - \frac{1}{2}H°_{O_2}$) and entropy ($\bar{S}_O - \frac{1}{2}S°_{O_2}$) of solution of oxygen through the stability field of wüstite whatever the temperature.
Transitional jumps are at 820 °C (W'i) and 1000 °C (Wi).
Dashed lines are drawn in extrapolated or metastable zones.**

**Isotherm at 1000 °C**: the three Wi are represented by successive linear segments. For the two transitions Wi → Wj at decreasing iron content, a jump is observed toward the *lower* molar heat and entropy of solution which are exothermic and exentropic. For $x_{1/2}$= 1.078: $\delta(\Delta \bar{H}_O)$ = -3.5 kJ, $\delta(\Delta \bar{S}_O)$ = -2.3 J.K$^{-1}$, and $x_{2/3}$ = 1.1154: $\delta(\Delta \bar{H}_O)$ = -10.5 kJ, $\delta(\Delta \bar{S}_O)$ = -8.3 J.K$^{-1}$.

Above 1184 K, the exothermic $\Delta \bar{H}_O$ for W1 increases with $x$ and $z$ corresponding to oxygen dissolution. For each $\underline{O}$ a vacancy and one or more electronic holes are created (*See below* ½ O$_2$ ⇄ $\underline{O}$ + V$_{Fe}^{q(')}$ + q h°, eq.(11)) simultaneously as the clustering progresses. For W3, $\Delta \bar{H}_O$ is decreasing with $x$, probably because of the long-range ordering. For W2, $\Delta \bar{H}_O$ varies in a manner intermediate between that for W1 and W3, which corresponds to the observation by (69Fen) that W2 would be a solid solution of W1 and W3.

Below 1184 K, the variations are not linear because the slight curvature of the isotherms $l' = f(x)$ was expressed using a parabolic term (*See* (79- 86- 89Val)). The variations are weaker for the three W'j. Nevertheless, an approximate convergence of parabolas $\Delta \bar{H}_O(x)$ is observed for $x$ = 1.0852 or $y$ = 1-$z$ = 0.9215 (±0.004), and is not defined as precisely for W' as it is for W. The convergence occurs for a composition different ($\delta y$ =1.3 10$^{-2}$) from that of point C ($y$ = 0.935).

**Isotherm at 820 °C**: the three W'i are represented by parabolic curves. Jumps (weaker in amplitude than the ones at 1000 °C) are observed toward higher values for the molar exothermic transitions W'j→W'j+1, $x_{1'/2'}$ = 1.071: $\delta(\Delta \bar{H}_O)$ = +10.9 kJ, $\delta(\Delta \bar{S}_O)$ = +6.3 J.K$^{-1}$ and $x_{2'/3'}$ = 1.1154: $\delta(\Delta \bar{H}_O)$ = +2.9 kJ, $\delta(\Delta \bar{S}_O)$ = +2.7 J.K$^{-1}$.

The Gibbs - Duhem relation leads to the iron solution property $\Delta \bar{H}_{Fe}$ from $\Delta \bar{H}_O$ (Fig. II-7). In the domain for W, exothermic transitional jumps are observed toward the higher values for the two transitions Wi → Wi+1 at equilibrium. At 1000 °C, the transition W1/W2 ($x_{1/2}$=1.0782) is characterized by $\delta(\Delta \bar{H}_{Fe})_{1/2}$ = 3.25 kJ.mol$^{-1}$, $\delta(\Delta \bar{S}_{Fe})_{1/2}$ = 2.56 J.K$^{-1}$.mol$^{-1}$. The transition W2/W3 ($x_{2/3}$=1.1164) is characterized by $\delta(\Delta \bar{H}_{Fe})_{2/3}$ = 12.06 kJ.mol$^{-1}$, $\delta(\Delta \bar{S}_{Fe})_{2/3}$ = 9.47 J.K$^{-1}$.mol$^{-1}$. In the domain for W', the equivalent jumps are exothermic as shown below in Fig. II-7 drawn for the transitions at 820°C.

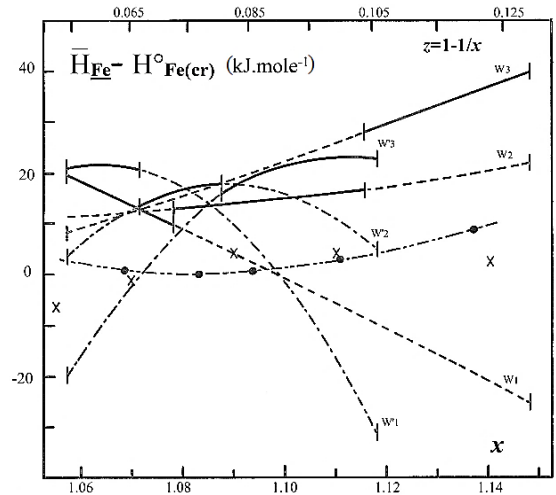

**Fig. II-7- Variations of the partial molar enthalpy ($\bar{H}_{Fe}$ - $H°_{Fe}$)$_{i,j}$ vs $x$ for the isotherms at 1000 °C (Wi) and 820 °C (W'j). For comparison: •---•: $\Delta\bar{H}_{Fe}$ from e.m.f. measurements at 1000 °C by Asao *et al.* [(70Asa) Table 3 p. 75]; ×: Darken and Gurry determinations [(45Dar) Table X p. 1408 )].**

As iron deficiency increases, the partial molar enthalpy of solution of iron $\Delta\bar{H}_{Fe}$ decreases for W1 and increases for W3, variations that are as opposite to that of $\Delta\bar{H}_O$. The five determinations of Asao (70Asa) at 1000°C are distributed on a curve noticeably parallel to the set of the successive segments concerning W1, W2 and W3.

The three '*varieties*' Wi of solid wüstite were defined for the first time in (63Val). Equations $x_{i/i+1}(T)$ of the lines -the so-called boundaries- separating the domain of stability of the single phases Wi and Wi+1 were then adjusted. Their analytical equations are:

$x_{i/i+1} = [(c_{i+1} - c_i) + T·(d_{i+1} - d_i)] / [(a_i - a_{i+1}) + T·(b_i - b_{i+1})]$

Numerically, they are

$x_{1/2} = (60412.0 - 41.3182T) / (56322.3 - 38.5509T)$  (II-11)

and

$x_{2/3} = (25297.2 - 18.310T) / (23670.0 - 17.1941T)$  (II-12)

A metastable boundary can also be envisaged:

$x_{1/3} = (85709.2 - 59.6282T) / (79992.3 - 55.7450T)$  (II-13)

The authors in (63Val) remarked that « *the corresponding terms in the numerator and the denominator* [of equations (II-11), (II-12) and (II-13)] *are not far from proportionality following the same ratio, close to 1.07* »: $(c_{i+1} - c_i) / (a_i - a_{i+1}) =$ $(d_{i+1} - d_i) / (b_i - b_{i+1}) \cong 1.07$, also close to Chaudron's point composition. It results that « *the three oxygen pressures $p'_1$, $p'_2$ and $p'_3$ at equilibrium with W1, W2 and W3, respectively, are nearly identical for $x \cong 1.07$* » at a given temperature (*e.g.*, p′ $\cong$ 8 10$^{-17}$, 4 10$^{-15}$, 8 10$^{-11}$ atm at 1184, 1273, or 1573 K. respectively), irrespective of i being. This result constitutes the reason why the property $\Delta\bar{H}_O$ (= -274.8 kJ.mol$^{-1}$) of equation (II-9) is shared by the three Wi's with the same composition $x \cong 1.07$ under the same equilibrium oxygen potential. Also, it explains why the oxygen partial molar properties vs $x$ converge at point $\Omega$. The iron partial molar properties also converge for $\Delta\bar{H}_{Fe} = +87.8$ kJ.mol$^{-1}$ at 1000°C.

At invariant point C (α-Fe, W'1, Fe3O4) of abscissa $x_C = 1.070_0$, these properties are merged ($\Delta\bar{H}_O = -290.4$ kJ.mol$^{-1}$, $\Delta\bar{H}_{Fe} = +21.4$ kJ.mol$^{-1}$ (86Val).

It is noticeable that a common *kinetical origin* ω is characterized by $x = 1.070 \pm 0.001$ in W' (*See* Annex 2 « Kinetics at 850 °C» by (10Wor)).

**4 - Triple point (W1, W2, W3) out of equilibrium**

« Point 8 (W1, W2, W3) » (*See* (64Val) Table III and Figure p. 3681) was determined as the intersection of the extrapolated boundaries W1/W2 and W2/W3 at temperature 302 °C and composition $y = 0.9322$, directly below point $\Omega$ and point C with $y_C \cong 0.935_0$. Manenc *et al.* observed a disproportionation near 300 °C in a miscibility gap by inducing a slow drop in temperature (64Her, 68Man). Fender and Riley remarked about wII that « *This behavior* [of $\partial(\Delta \bar{G}_O$ vs 1-x$)/\partial T$] *in wII region is characteristic of a two-phase system above its critical temperature (…) Extrapolation into a metastable $Fe_{1-x}O$ phase indicates that phase separation into wI and wIII* [not explicitly demonstrated] *would occur near 315 °C* » (69Fen).

As early as 1977, **Andersson and Sletnes** (77And) studied conjointly the ordering and the preceding behavior below 300 °C in reference with the spinodal decomposition (71Mor) using electron diffraction and microscopy (dark field images). Typical patterns were attributed to the phase P′ (periodicity of $2.6a_o$), in the composition interval $y$ (of $Fe_yO$) ∈ [0.95-0.92]. After heat treatment for 20 min at 300 °C, a tetragonal or orthorhombic symmetry was observed « *also from the shape of the ordered P′-type regions* ». For compositions less than $y = 0.91$ (in $Fe_{0.91}O$), the more ordered structure identified as P″ was observed even after quenching. Its parameters are $a=b=c=5a_o$.

These different observations indicate that W2 can be represented as a solid solution of W1 and W3 (*See* (R-10Des) p. 91-96), consistent with the determinations of $[\Delta\bar{H}_{\underline{Fe}}(x)]_\mathbf{i}^\mathbf{e}$ below.

## 5 -Trends to ordering in the sub-phases for increasing $z$

The difference $[\Delta\bar{H}_{\underline{Fe}}(x)]_\mathbf{i}^\mathbf{e}$ between the partial molar property of iron in $(FeO_x)_\mathbf{i}$ and $FeO_{xo(i)}$ on the boundary with Fe was envisaged as an « *excess type* » property in order to tentatively establish links between thermodynamic and structural properties (65Kle).

The corresponding formalism previously established for $[\bar{H}_{\underline{O}}(x)]_\mathbf{i}, [\bar{H}_{\underline{Fe}}(x)]_\mathbf{i}, [\bar{S}_{\underline{O}}(x)]_\mathbf{i}$ and $[\bar{S}_{\underline{Fe}}(x)]_\mathbf{i}$ leading to Figs. II-6 and II-7 above is simply:

$[\Delta\bar{H}_{\underline{O}}(x)]_\mathbf{i}^\mathbf{e} =$
$[\bar{H}_{\underline{O}}(x) - \bar{H}_{\underline{O}}(x)]_\mathbf{i} = 19.14475(a_\mathbf{i}/2(x-x_{o(\mathbf{i})}))$ (II-14)

$[\Delta\bar{S}_{\underline{O}}(x)]_\mathbf{i}^\mathbf{e} =$
$[\bar{S}_{\underline{O}}(x) - \bar{S}_{\underline{O}}(x_o)]_\mathbf{i} = -19.14475(b_\mathbf{i}/2(x-x_{o(\mathbf{i})}))$ (II-15)

$[\Delta\bar{H}_{\underline{Fe}}(x)]_\mathbf{i}^\mathbf{e} =$
$[\bar{H}_{\underline{Fe}}(x) - \bar{H}_{\underline{Fe}}(x_o)]_\mathbf{i} = -19.14475(a_\mathbf{i}/4(x^2 - x_{o(\mathbf{i})}^2))$ (II-16)

$[\Delta\bar{S}_{\underline{Fe}}(x)]_\mathbf{i}^\mathbf{e} =$
$[\bar{S}_{\underline{Fe}}(x) - \bar{S}_{\underline{Fe}}(x_o)]_\mathbf{i} = 19.14475(b_\mathbf{i}/4(x^2 - x_{o(\mathbf{i})}^2))$ (II-17)

Substituting the coefficients $a_\mathbf{i}$, $b_\mathbf{i}$ and $x_{o(\mathbf{i})}$ on the γ-Fe/$W_{o(\mathbf{i})}$ boundary by their numerical values at 1273 K yields Table II-1 with values in J.mol$^{-1}$ and J.K$^{-1}$.mol$^{-1}$ for $\bar{H}$ and $\bar{S}$, respectively. The term $[\Delta\bar{H}_{\underline{Fe}}(y)]_\mathbf{i}^\mathbf{e}$ is determined from the corresponding term for oxygen $\underline{O}$ by means of the Gibbs-Duhem relation. Its sign as well as the sign of its derivative $\partial/\partial x$ is the sign of -$a_\mathbf{i}$. Therefore, this term corresponds to the energetic effect δU on the iron lattice for the interaction between ½ $O_2$ and a mole of $FeO_x$ to which it is added, at constant iron content. For each additional $\underline{O}$, a vacancy and zero, one or two electronic hole(s) are created as:

½ $O_2$ ↔ $\underline{O}$ + $V_{\underline{Fe}}^{\mathbf{q}(\prime)}$ + q h° (q=0,1,2) (II-18)

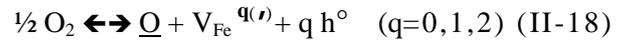

The values of $[\Delta\bar{H}_{\underline{Fe}}]_\mathbf{i}^\mathbf{e}$ (Table II-1) show that:

1. for W1, the internal energy $U_m$ of the iron lattice (the term $P\Delta V_m$ is negligible compared to the contribution $\Delta U_m$ in $\Delta H_m$) decreases as $z$ increases, which corresponds to the exothermic formation energy of clusters (m:n) that are thus favored (*See* Gokcen (R-75Gok) p. 241-44, (R-86Gok) p. 43-49, Desré and Hodaj (R-10Des) p. 57-58, 76-83)).

2. conversely, W3 is characterized by an increase in the lattice energy, which corresponds to the completion of the cluster percolation and the formation of the superstructure (2.55a x 2.55a x 2.55a) named P″ by Manenc (63-68Man) and characterized using HREM by (74Iij);

3. likely because W2 is a solid solution of W1 and W3, the variation of $U_m$ is intermediate (*See* (69Fen) and in Figs. II-6 and 7).

Such trends were discussed previously by Fender and Riley (*See* (69Fen) p. 797) when considering $\Delta\bar{G}_{\underline{O}}$ and $\Delta\bar{H}_{\underline{O}}$.

**TABLE II-1 - Variations at 1273 K of "excess properties" (relations (II-14 to 17))**

| $FeO_x/Fe_{1-z}O$ | $x$=1.0950 or $z$=0.0868 | | $x$=1.1420 or $z$=0.1243 | |
|---|---|---|---|---|
| W1: $a_1$= 46753.4<br>$b_1$= -7.3781<br>$x_{o(1)}$= 1.0572*<br>$z_{o(1)}$= 0.0541 | $[\Delta\bar{H}_O]_1^e$<br>16.90<br>$[\Delta\bar{S}_O]_1^e$<br>2.67 | $[\Delta\bar{H}_{Fe}]_1^e$**<br>-18.19<br>$[\Delta\bar{S}_{Fe}]_1^e$**<br>-2.87 | $[\Delta\bar{H}_O]_1^e$<br>37.94<br>$[\Delta\bar{S}_O]_1^e$<br>5.99 | $[\Delta\bar{H}_{Fe}]_1^e$<br>-41.72<br>$[\Delta\bar{S}_{Fe}]_1^e$<br>-6.58 |
| W2: $a_2$= -9568.9<br>$b_2$= 31.1728<br>$x_{o(2)}$= 1.0522<br>$z_{o(2)}$=0.0496 | $[\Delta\bar{H}_O]_2^e$<br>-3.92<br>$[\Delta\bar{S}_O]_2^e$<br>-12.78 | $[\Delta\bar{H}_{Fe}]_2^e$<br>4.21<br>$[\Delta\bar{S}_{Fe}]_2^e$<br>13.76 | $[\Delta\bar{H}_O]_2^e$<br>-8.23<br>$[\Delta\bar{S}_O]_2^e$<br>-26.80 | $[\Delta\bar{H}_{Fe}]_2^e$<br>9.03<br>$[\Delta\bar{S}_{Fe}]_2^e$<br>29.41 |
| W3: $a_3$= -33238.9<br>$b_3$= 48.3669<br>$x_{o(3)}$=1.0482<br>$z_{o(3)}$=0.0460 | $[\Delta\bar{H}_O]_3^e$<br>-14.90<br>$[\Delta\bar{S}_O]_3^e$<br>-21.68 | $[\Delta\bar{H}_{Fe}]_3^e$<br>15.97<br>$[\Delta\bar{S}_{Fe}]_3^e$<br>23.24 | $[\Delta\bar{H}_O]_3^e$<br>-29.85<br>$[\Delta\bar{S}_O]_3^e$<br>-43.44 | $[\Delta\bar{H}_{Fe}]_3^e$<br>32.69<br>$[\Delta\bar{S}_{Fe}]_3^e$<br>47.57 |

\* For $x$= $x_{o(i)}$, all terms are equal to zero. ** The terms $[\Delta\bar{H}_O]_i^e$ and $[\Delta\bar{H}_{Fe}]_i^e$ are expressed in kJ.mol$^{-1}$, and the terms $[\Delta\bar{S}_O]_i^e$ and $[\Delta\bar{S}_{Fe}]_i^e$ in J.K$^{-1}$.mol$^{-1}$.

The positive and negative values of the term $[\Delta\bar{S}_O]_i^e$ correspond mainly to a vibrational entropy change (R-00Smy) in the oxygen lattice, without transpositional mixing effect (69Lyk). In the case of the initially disordered pseudo-phase W1, the excess entropy term weakly increasing with $z$ shows that the disorder due to the formation of point defects is predominant. Reversely in the case of the pseudo-phase W3, a strong decreasing with $z$ corresponds to an increased ordering at short and long range. These entropy variations are related to the increasing frequency of the Fe-O bond because of larger clusters and/or an increasing degree in their percolation until completion and superstructure. The pseudo-phase W2 can be thought to appear at the starting of the percolation of the clusters, when the lattice becomes more rigid.

**6 - Burgmann** (R-75Bur) gathered some experimental results (electrical conductivity and transition p-to-n). Because of numerous analogies with pyrrhotine $Fe_{1-z}S$ (68Bur), he described tentatively the defect structure in terms of clusters (*See* (R-75Bur) Fig. 5 p. 173). He proposed a conduction model involving the formation of an acceptor band or of new states produced by the interaction of defects as $z$ increases. The lattice contraction reduces the gap between the valence and conduction bands, inducing a p-to-n transition for compositions $z$= 1/12 = 0.083 at 900 °C and $z$ = 1/13 = 0.076 at 1300 °C (*See* (R-75Bur) Fig. 5 p. 173, and Hillegas and Wagner jr (68Hil)). Such an overlap was also considered later by Molenda *et al.* (87Mol**1,2**).

**7 - Manenc's phases P, P′, P″**

Using X-ray diffraction, J. Manenc (63- 68Man) characterizes the phases after (rapid) quenching into water. He names the phases P and P′. For an iron rich wüstite ($z$ < 0.08), vacancies and ions $Fe^{3+}$ seem to be randomly distributed. This corresponds to the phase P which has a weakly incomplete NaCl-type structure alone. Other diffraction patterns of more iron deficient samples (0.08< $z$ < 0.11) have additional reflections corresponding to a cubic cell with a repeating distance 2.6$a_o$. Diffuse trails parallel to <100> directions connect these reflections. These features are characteristic of the phase P′ alloyed with the phase P, and individually observed only in « oxygen rich » samples. Iron vacancies are ordered in the two phases.

Diffraction for a heated single crystal such as $z = 0.08$ shows that a superstructure exists in the range 800-1000 °C for this composition. The phase P″ is observed in wüstite such as $z \cong 0.10$ near 300 °C and when slowly cooled, at the threshold of disproportionation yielding Fe and Fe3O4. New superstructure lines are then characteristic of a commensurate 5X cubic cell.

**8 - Greenwood and Howe**

Mössbauer spectra were interpreted in modeling the asymmetrical doublet as evidenced of « *slightly differing defect structures anticipated by Vallet and Raccah* (65Val), *and Fender and Riley* (69Fen) *at high temperature* » (72Gre). The quenching process produces nuclei leading to Manenc phases P, P′, P″ [(P+P′) for $z \in$[0.05-0.08], (P′) for $z \in$[0.08-0.10], and (P″) for $z > 0.10$] below point C. The phases are detectable from fine variations of quadrupole splitting resulting from cubic symmetry distortions. These distortions can be evaluated for « single cluster » (4:1) and Koch and Cohen « four-fold cluster (13:4) ». From a sample such as $z \cong 0.05$, the disproportionation gives rise to $Fe_{0.918}O$ (defect rich phase P′) and $Fe_{0.976}O$ (defect poor phase P), because of a mixture of the (13:4) and (4:1) clusters.

Among numerous Mössbauer studies reviewed by Long and Grandjean (R-91Lon), Checherskaya *et al.* (72Rom) (73Che), Hrynkiewicz *et al.* (72Hry), Pattek-Janczyk *et al.* (86Pat) are of particular interest. These authors modeled the asymmetric doublet with « a singlet » attributed to $Fe^{3+}$ in tetrahedral sites, and two « crossed doublets » attributed to octahedral $Fe^{2+}$ and $Fe^{3+}$. They associated the maximum observed in the variation of the quadrupole splitting to the p-n transition described by (62Tan). This transition is likely the same as the transition « metal ↔ Mott insulator » identified by Molenda *et al.* (87Mol**1,2**).

**9 - Goodenough's** modeling was proposed as early as 1971. Considering the relations $z \propto (p')^{1/s}$ (61Smy) associated with thermogravimetric analyses, and $\sigma \propto (p')^{1/s}$ corresponding to the conductivity (66Gei), the result s=6 in the W1 subdomain above 1060 °C would characterize a statistical distribution of vacancies and electronic holes ($Fe^{3+}$), possibly corresponding to Manenc's phase P (68Man). Below 1060 °C, the changes in values of s with 4< s <6 would correspond to an increasing association between vacancies and holes. In the W2 domain, triplets $Fe^{3+}$ - $V_{Fe}$ - $Fe^{3+}$ are partially ordered, improving Madelung energy. Numerous (4:1) clusters which form a superstructure according to Koch & Cohen (69Koc) correspond to W3. A mapping of the transfers of iron ions, and electronic exchanges is proposed as a mechanism, needing *small polarons* related to energy differences between the top of the valence band and an acceptor level (R-71Goo). Later, this author with Gleitzer will consider the wüstite from the point of view of its electronic properties and crystallographic structure (R-85Glei).

**10 - Zvintchuk's team**

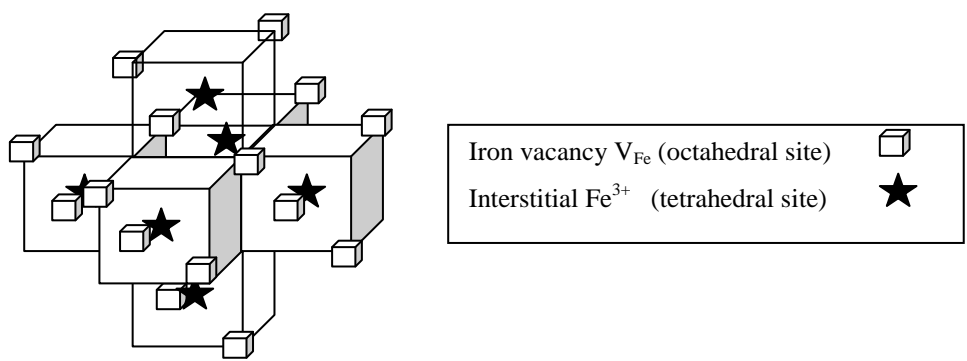

**Fig. II-8 – The m3m cluster (16:6) of <110> type corresponding to $(z+t)/t = 2.66$ from Tchiong Tki Khong *et al.* (73Tch).**

Tchiong Tki Khong in Zvintchuk's team (73Tch), in collaboration with J. Manenc, studied samples of wüstite of composition $z \in [0.065\text{-}0.108]$ quenched from 1100°C. Electron diffraction revealed three « forms I, II and III » corresponding to the modifications wI, wII and wIII of the phase diagram by Fender and Riley (69Fen). Electron diffraction revealed three « forms I, II and III » corresponding to the modifications wI, wII and wIII of the phase diagram by Fender and Riley (69Fen). They described a (16:6) defect complex of point symmetry m3m, formed of six tetrahedral corner sharing clusters (4:1), distributed on the six faces of an empty cube. This cluster is represented on Figure II-8 above. An « obvious correspondence » of phases P, P′, P″ with the three Wi is specified. Later work by Shaïovitch *et al.* (87Sha) describes a superstructure with a quadratic cell P$\bar{4}$3m is for a wüstite $Fe_{1-0.10}O$.

## 11 - Kinetic transitions

More than 40 years after Vallet *et al.* (63- 64- 65Val), Wagner jr *et al.* (66Gei) and Fender and Riley (69Fen), **Worral and Coley** (10Wor) evidenced kinetic transitions related in an obvious manner to the pseudo phases. Following the process first used by Temkin *et al.* at the surface of iron oxides (59Str) (61Tem), and then applied by Grabke (65Gra), Belton *et al.* (81Cra), Sano *et al.* (96Mor), later Zhang *et al.* (12Hu) (13Zha), Worral and Coley studied the kinetics of the change in the equilibrium of CO2/CO mixtures in which a controlled proportion of $^{13}CO_2$ species was introduced. In such an environment, a specific equilibrium of the carbon isotopes is established following the exchange equation:

$$^{13}CO_2 + CO \leftrightarrows {}^{13}CO + CO_2$$

which can be viewed as the sum of two half - reactions

$$^{13}CO_2 \rightarrow {}^{13}CO + O_{ads} \quad \text{and} \quad CO + O_{ads} \rightarrow CO_2$$

as reflecting oxygen exchange on the surface, and being conditioned by the electronic structure in the bulk. The apparent rate constant is expressed as a function of oxygen activity as: 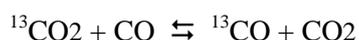 $k_a = k_0 . a_O^{-m}$

or 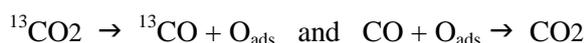 $\log k_a = - m.\log a_O + \log k_0$.

The variation of log $k_a$ with the activity $a_O$ defined as being the ratio p(CO2)/p(CO), can be described by three success-sive linear segments at 995 °C, with m = 0.51 (w1), 0.66 (w2), 1.03 (w3) in ((10Wor), Fig. 8 and equations [36]-[38]).

The authors observed only one transition attributed to two pseudo phases w'1-2 and w'3 along the isotherm at 850°C, and no change was observed in the vicinity of the forecasted boundary w'1/w'2 by (69Val). The only observed transition appears as a discontinuity at the location of the transition W'2 ↔ W'3 forecasted for $x = 1.090$ at equilibrium. Nevertheless, they stated that three pseudo-phases w'1, w'2 and w'3 probably exist, but that two of them could not be separated, either because they were not detectable by electrical conductivity measurements (*See* (66Gei)), or because they were possibly '*obscured by the scatter*'. Finally, Worral and Coley declared that, in their study, transitions at 850 °C and 995 °C were observed (except for W'1/W'2) '*correspond(ing) exactly with the proposed pseudo-phase boundaries*' in quite good agreement with the phase diagram by Vallet and Carel (89Val).

Because of the identity of coefficients m in the correlations concerning Wi and W'j for i= j, they attributed in a following paper (R-13Wor) three predominant clusters (7:2) type II to w1 and w'1, (12:4) type II to w2 and w'2 and (16:5) type I to w3 and w'3.

NOTA: In Annex 2 and Fig. A2-2, a graphical and numerical re-analysis of Fig. 9 by ((10Wor) p. 821) shows a possible separation of the data in three sets at 850 °C, which can likely correspond to the three W'j, provided that some data or groups of data are located in subdomains not proposed as stable.

## 12 - Takayama and Kimizuka's phase modifications

Using thermogravimetric analyses under equili-brium oxygen pressure p′ defined by CO2/H2 mixtures, Takayama and Kimizuka (80Tak) observed significant modifications in curves representing $l'$ *vs* log $x$ in FeO$x$. The authors concluded to the existence of a transition between

« pseudo-phases » (*See* (80Tak) Table I and Annex 1 d-). Their conclusion is in agreement with the results from Bransky and Hed in the same domain of temperature (68Bra) (*See also* Annex 1 b-), but disagrees with results of Fender and Vallet, and particularly with the results from Vallet as they do not show a break in the isotherms above 911 °C.

Typical changes in slopes are clearly observed from their data. To illustrate their results, we have transformed their representations for one temperature T=1523 K. For this temperature, their data were interpreted in terms of two relations only:

- For low $x$ values ($x = 1/(1-z)$):

$l' = M_1.x + B_1 = 28.512\ x - 41.366$

- For high $x$ values:

$l' = M_2.x + B_2 = 23.288\ x - 35.715$

The new representations for $l'$ vs log $z$ are done in Fig. II.9. Two coefficients s can be defined as a first step, using the hypothesis of a transition implying two pseudo-phases proposed by authors (80Tak): s close to 4 and s close to 7.5. However, three correlations can be defined as follows:

$l' = 4.4539\ \log z - 5.5685$, $l' = 6.1378\ \log z - 3.6876$, $l' = 9.5059\ \log z - 0.5046$.

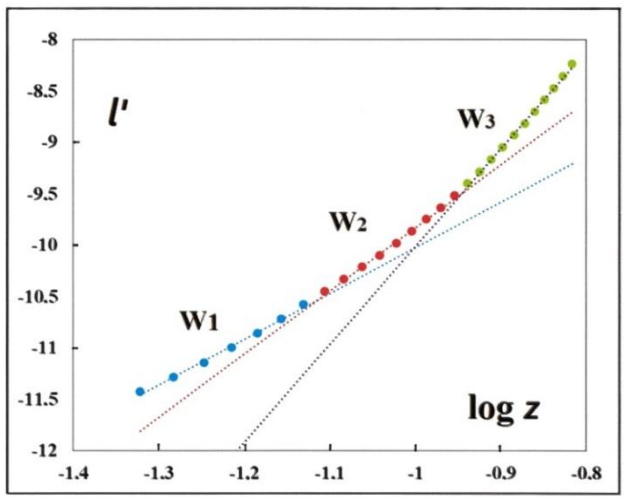

Fig. II-9 - Variation of $l'$ vs log $z$ for $Fe_{1-z}O$ at 1250 °C. Three sets of data can be defined as three linear relations.

From these relationships, it is possible to derive three values for s. A first domain for low $z$ values yield a coefficient s = 4.5. A second domain appears as being nonlinear, with an average slope corresponding to s = 7.5. However, due to the curvature of the representation it is possible to interpret this nonlinear domain in terms of two linear sequences with two coefficients s = 6.1 and s = 9.5. These two coefficients are quite close to those obtained from the data in Vallet *et al.*

### 13 - Electrical transitions. Percolation.

In an extensive review about electrical properties, **Gleitzer** (R-97Gle) points out that the electric conductivity of wüstite « *cannot be understood without percolation models which take into account the cluster populations as function of temperature and oxygen activity* ».

The superimposition of the two kinds of p-n and $W_i/W_{i+1}$ transitions, almost independent of each other, and both function of $z$, was referenced in the review by Burgmann (R-75Bur).

Because of this common stress, the characterization of the p-n transition and that of the electronic structure are not easy to correlate. Since the percolation process is characterized experimentally by a brief steep variation of the studied property at a threshold composition, the curves for the electrical conductivity *vs* composition (63Ari) (66Gei) (67Koz) can be interpreted in terms of upgradable clusters and long-range ordering in the subdomains of $W_1$, $W_2$ and $W_3$.

Later in this review (*See* Section IV, Fig. IV-1), a statistical distribution of clusters with two types of percolation modes is schematically represented through the phase diagram. This cluster distribution can justify structurally the singularities of the integral and partial thermodynamic properties (*See above* Figs. II-1 -2 -6 -7 -9). Also, it can justify the distinct rate constants observed in kinetic processes (*See above* Section II-7 § 11 Kinetic transitions (10Wor) and Annex 2: Figs. A2-1 and -2).

**The p-n transition**

**Tannhauser** (62Tan) observed a thermoelectronic p to n transition around 1300 °C by thermal e.m.f. analyses of a highly nonstoichiometric wüstite. More precisely with Bransky (67Bra), a change in sign of the Seebeck effect is observed for polycrystalline samples near O/Fe ~1.09 or $z$ ~ 0.083 at T ∈ [1010-1310°C] and near $z$ ~ 0.066 at 910 °C (Fig. II-10). The authors formalized the Seebeck coefficient α following (63Hei) in the case of completely ionized vacancies when the « hopping process » is predominant.

$$\alpha = (k/e)(A/kT + \ln (c_o-c)/c)$$

where c = 'concentration' of charge carriers, $c_o$= 'concentration' of available sites for the charge carriers.

The same formalism was used by **Lafollet and Duquesnoy** (77Laf) in the case of non-localized charge carriers, and when the wüstite is considered as an intrinsic semiconductor, *i.e.,* doped by its own impurities. They modeled α = 0 for $z$ = 0.104 at 1100 °C, and $z$ = 0.108 at 1000 °C (*See* Fig. II-10).

Like in previous studies, Hodge and Bowen (81 Hod)

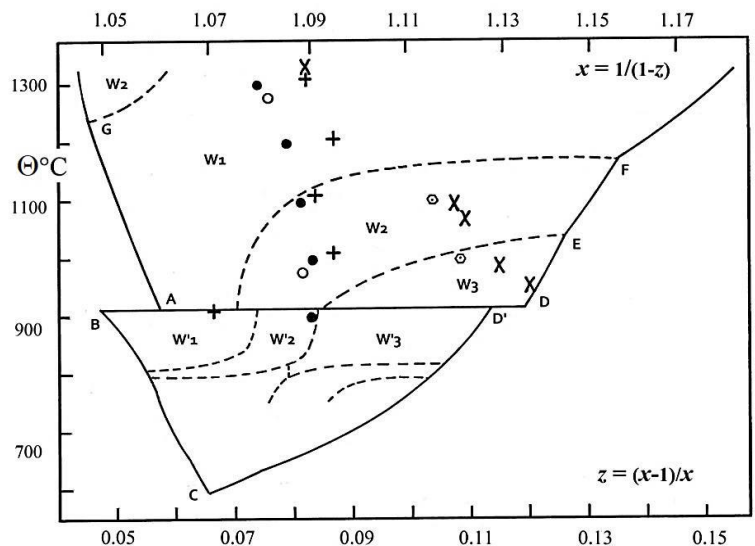

Fig. II-10 - Some determinations of the p-n transition by thermal e.m.f..
✗ : G. H. Geiger *et al.* (66Gei) not corrected for platinum ; ✚ : I. Bransky and D. S.Tannhauser (67Bra) ; ○: W. J. Hillegas jr and J. B. Wagner jr (67Hil) ; ⊙ : P. Lafollet and A. Duquesnoy (77Laf) by modeling; ● : E. Gartstein and T. O. Mason (82Gar) by re-examination of W. J. Hillegas jr's thesis (1968).

stated that the experimental sign change from positive to negative of the thermal electromotive force occurred at increasing values of $z$. Their model considered principally cluster (4:1) and electron holes trapped in octahedral sites adjacent to cluster vacancies. The thermally activated hopping of these holes permits continuous paths through the crystal because of the high point defect concentration between the zones separating the clusters.

In their studies on electrical conductivity, **Ariya and Bratch** (63Ari) and **Kozheurov and Mikhailov** (67Koz) observed modifications to the relationship $\sigma \propto (p')^{1/n}$, closely related to the relationship $z \propto (p')^{1/n}$, where their coefficient n (≡ s) varied in such a way that three subdomains could be defined. They found an increase of the conductivity in two increases from $z \cong 0.048$ to 0.079, and then to 0.097, corresponding to the Fe/W and W/Fe3O4 boundaries.

**Neuschütz** and **Towidi** (70Neu) established a variation of the resistivity $R^{-1}(T, x)$ for $T \in [700-1220\ °C]$ from which they infered that the defect electrons jump between $Fe^{2+}$ and $Fe^{3+}$, the activation energy of the process being 9.6 kJ.mol$^{-1}$.

Following **Molenda** *et al.* (87Mol**1,2**), the conductivity should be the sum of an « extrinsic conductivity » $\sigma_y$ related to the nonstoichiometry and of an intrinsic component $\sigma_o$ disregarded until now, independent of the defect concentration, *i.e.*, of $p'$. Charge transport at high temperature occurs simultaneously in the valence and conduction bands, and the dopant acceptor band due to $Fe^{3+}$ in octahedral and tetrahedral sites. Near the composition $z \approx 0.08$, the acceptor band overlaps with the valence band giving rise to a broad metallic band close to the thermoelectric transition « p to n » for $z \approx 0.09$, and corresponding to the stability limits of W1 and W2-3. Experimental details can be found in ((87Mol) Fig.2 p. 519 and Fig. 4; *See also* Annex1 e-, where successive segments corresponding to the Raccah-Vallet's isotherms are observable).

More recently, **Toroker** and **Carter** (12Tor) examined the possible improving of the conductivity of Fe$_{1-z}$O by doping the p- or n-conduction mode with the aim of stabilizing the nanoscale fabrication of wüstite, and then adapting the wüstite nano-particles to create efficient solar light conversion materials. They list the procedures allowing the calculation of Marcus theory parameters for « clusters models » with hydrogen dopant and substitutional Cu$^+$, Li$^+$ and Na$^+$. They conclude that « *iron vacancies drastically limit the hole conductivity* » while Li, H and Na dopants improve it. So, it would be interesting to suppress the forma-tion of vacancies by « alloying FeO », which would amplify the p-type conductivity.

Nota: An extensive thermodynamic study of multiple substituted wüstites can be found in the work of Lykasov *et al.* which is reviewed partly in (R-87Lyk) (99Kat). Structural aspect of calcio- and magnesio-wüstites is available *infra* (Table III-1, and in Fig. III-4), (90Car).

**14 - Other kinetical and diffusional works.**

**Desmarescaux and Lacombe** (63Des) studied the self-diffusion of iron using isotopes $^{55}$Fe and $^{59}$Fe. They stated that D$_{Fe}$ increases with $z$ at a given temperature T, and decreases with T at fixed $z$. They deducted that self-diffusion proceeded through a vacancy driven mechanism, especially noticeable at high temperature. At lower temperatures, fewer free vacancies and more complex vacancy-electron holes exist. The number of vacancies associated to $Fe^{3+}$ cations (the clusters) increases with $z$.

**Landler and Komarek** (66Lan) measured the isothermal kinetics of mass loss *vs* time ΔM(t) in H2/H2O mixtures by reduction of polycrystalline samples by thermogravimetry. In their calculations, coefficients of iron self- and chemical diffusion decrease for increasing $z$ for $T \in [800 - 1050\ °C]$ because of « *the increased freezing due to repulsive interaction of the vacancies* ». This observation corresponds primarily to the formation of clusters, then to the ordering of clusters in the bulk (*See* Excess molar partial properties p. 13-14). Considering the relationship log(ΔM/ΔM$_o$) = nlog t + k in which ΔM$_o$ is the total mass loss, reduction kinetics data from the authors at 950, 1000 and 1050 °C can be separated into three regimes II, III, and IV (*See* Annex 2, Fig. A2-1). At 1000 °C, the calculated mean values of n are -0.72 (W3?), -0.49 (W2?) and -0.3 (W1?), , respectively (67Car). A brief anharmonic sequence is observed at the transition between reduction processes. The increase of 0.6 eV in the activation energies E$_{III}$ = 1.31 eV and E$_{II}$ = 1.89 eV for the reduction sequences of W3 and W2 corresponds to the change in ordering process observable on the boundary W3/W2 for decreasing values of $z$. The three coefficients n are primarily the signature of the bulk modifications (66Lan) while the three kinetic coefficients m in (10Wor) are primarily characteristic of surface phenomena (*See* (81Now) bulk-surface differences by reduction).

Similarly, successive different reduction sequences using H2O/H2 gas mixtures that exhibit transitional anomalies can be observed (*See* Annex 1 c-) in **Rieke and Bohnenkamp** (66Rie).

**Rickert and Weppner** (74Ric) combine a solid galvanic cell for the ionic conduction by $O^{2-}$ through the sample while having the other sample face in contact with a platinum foil. The diffusion is studied *in situ* during the relaxation from a composition to another driven by controlled impulses of electrical potential. Their main result is that the chemical diffusion coefficient increases with increasing $z$ (*i.e.*, the vacancy concentration), which is in agreement with Landler and Komarek (66Lan) and not with any other authors. Transitional anomalies are clearly noticed (*See* (74Ric) Fig. 6 and 7 p. 1856, and Fig. 8 p. 1857).

## 15 - Relationships between defect structures and the phase diagram

We review key publications from authors who have attempted to solve the key challenge formulated by Carl Wagner in 1930 of linking more tightly defect structure and thermodynamics, such as partial molar properties or the equilibrium constant K of formation of the cluster. Kinetics (diffusional) relations were also developed.

**Per Kofstad and Hed** (68- R-72Kof) remarked that $x$ or $z$ increases at constant $p'$ as T decreases, which is in contrast with other oxides, and likely due to a complex defect structure. They emphasized that it is *doubtful* that all the iron vacancies are be doubly charged. The electron holes associated to a vacancy should jump into only one of the 12 next nearest octahedral neighbors ($Fe_{Fe}^\circ$) and one tetrahedral site ($Fe_i^{\circ\circ\circ}$). Most of octahedral vacancies would be singly charged. A model can thus be elaborated from the Roth complex $[V_{Fe}^{m(')} - Fe_{in}^{(\circ)} - V_{Fe}^{p(')}]^{q(')}$, n=2,3, (m,p) = 0,1,2. In the relationship $\sigma \propto p'^{(1/s)}$, the s value is larger than the value in the similar relationship connecting z to p' because vacancies are neutral.

**Toft Sørensen** in a book devoted to nonstoichiometric oxides (R-81Sor), and then with El Sayed Ali (85Sor), tentatively identified the defect structure of the « subphases ». They used thermogravimetric data [$l'$- $z$] at 1000 - 1300 °C by Bransky and Hed (68Bra) exclusively. They considered three main defects (doubly charged vacancy $V_{Fe}''$, and tetrahedral complex defects $(4:1)^{(m)-}$ and $(6:2)$) as defined by Catlow *et al.* (75-80Cat). They focused on the defect cluster differently ionized, which led them to define a linear plot of $l'$ *vs* log $y$ ($\equiv z$) with minimal slopes $1/s$ in the relation $z \propto (p')^{1/s}$. A second criterion concerns the so-called exclusion envelope, *i.e.,* the long-range order as a layer or a close packing structure. They localized in the wüstite equilibrium area T($z$) the predominant defect (4:1) plus a layer structure, (4:1) plus a close packing, (6:2) plus a layer structure, according to ΔT layers, above 1300 °C, between 1300 and 1200 °C, and below 1200°C, respectively. In addition, iron vacancies $V_{Fe}''$ are present on the right side of a line at composition $z$ close to 0.09 (*See* (85Sor) Fig. 7 p. 20). The pseudo-phases of Fender (69Fen) and Vallet (70Val) are thus « roughly depicted and structurally characterized ».

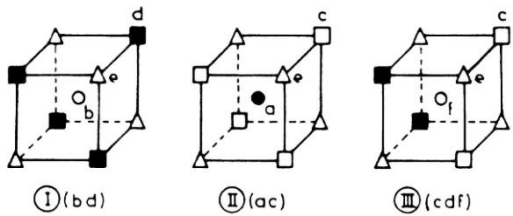

Fig. 3. Structural unit blocks S.S. ; ■ : (16d) positions, □ : (16c) positions, O : (8b) positions, ● : (8a) positions, ✱ : (48f) positions, △ : (32e) positions, (ref. 2, see Table 7.1. p. 186).

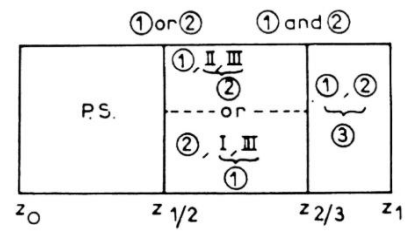

Fig. 4. Short range order in wüstite in terms of S.S. ① and C.C. ① in the three sub-domains of the wüstite homogeneity range. C.C. 1 : I-6III, C.C. 2 : II - 6III, C.C. 3 : I - 6III - II. In the subdomain ($z_0$, $z_{1/2}$) : II and III increase, I decreases ; in ($z_{1/2}$, $z_{2/3}$) : II and III decrease, 1 or 2 increases ; in ($z_{2/3}$, $z_1$) : 1 and 2 decrease, 3 increases.

*by* Men and Carel (83Men**1**)

**Fig. II-11** – The three pseudo phases from $z_o$ on the γ-Fe/W1 boundary to $z_1$ on the W3/Fe3O4 boundary described with the help of three cluster components ①, ②, ③.

In the course of their theoretical works about oxides, **Men** *et al.* (70Men) (R-77Men) were concerned with the wüstite. Men and Carel applied the C(luster) C(omponent) M(ethod) to model the wüstite solid solution (82Men) (83Men**1,2**) (85Men). The « cluster components » were defined from crystallographic positions forming unit blocks in the general spinel structure. For example, three cluster components ①, ② and ③, and their concentration as a function of $z$ allowed to build the lattice of the wüstite $Fe_{1-z}O$, and to model some of its physical properties, taking into account the pseudo-phases (Fig. II-11 above).

**Rekas and Mrowec** (87Rek) (*See also* (R-87Mro)) worked on defect clustering in the w<u>u</u>stite phase using a collection of data ($z$, log p′) from the literature (45- 46Dar) and (80Tak). They envisaged all the defect clusters suggested by Catlow and Fender (75Cat), *i.e.*, (4:1), (6:2), (8:3), (10:4), (12:4) and (16:5) with several ionization degrees including the highest. Elaborate calculations of the equilibrium constants led them to the assessment of the enthalpy and entropy of formation of all these ionized clusters, then to $\Delta \bar{H}_O$ independently of $z$ and T, and to $\Delta \bar{S}_O$ for $z \in [0.055-0.145]$. They concluded that the (4:1) cluster is the most likely to occur and that the ionization degree increased with temperature to reach the maximum value (5⁻) at 1573 K.

**Nowotny and Rekas** (89Now) tentatively described the « *defect structure and thermodynamic properties of the wustite phase $(Fe_{1-y}O)$* ». They investigated the clustering of point defects (iron vacancies and electron holes) across the entire domain using the Debye-Hückel theory, which allows to use a product $\gamma.[x_i]$ in place of $[x_i]$, particularly for i =h°. With this approach, the activity coefficients are confounded with the sole mean coefficient $f_\pm$ determined as a function of the ionic force of the solution (and not of its square root). They modeled the interactions during the formation of the cluster (4:1) taking into account activity instead of concentration. Sets of data ($z$, log p′) and ($\Delta \bar{H}_O$, $z$) were gathered from the literature. They evaluated the enthalpy and entropy of formation of a $(4:1)^{5-}$ cluster at -396.7 kJ.mol⁻¹ and 267.4 J.K⁻¹.mol⁻¹, respectively. They concluded that this cluster $(4:1)^{n-}$ would be the constitutive module of more extended clusters (89Now). An expression of the equilibrium constant for the cluster formation is determined across the whole domain independently of the existence of pseudo-phase transitions.

As soon as 1975, **Catlow and Fender** (75Cat) tried to access the defect structure by means of large scale computation of the binding energies following the Born model. They envisaged in a theoretical manner the clusters unity (4:1), edge-sharing (6:2) and (8:3), corner sharing (16:5) precursor of $Fe_3O_4$. The location (zone $z$, T of the phase diagram) of these clusters is suggested qualitatively, but not in relation with the pseudo phases already known by (69Fen). Catlow *et al.* (79Cat) assessed the formation energy and the relative stability of the clusters by means of the Mott-Littleton method. The cluster (12:4) <110> type by Lebreton and Hobbs (83Leb) is considered as being the most stable extended cluster after clusters (6:2) and (8:3). In a review paper (87Cat), four classes of clusters are defined:

1. corner-sharing (4:1), (7:2), and (16:5) clusters;
2. edge-sharing clusters (6:2), (8:3), (12:4) (83Leb);
3. ZnS blende like cluster (10:4) (86Grim) (87Gav);
4. cluster (13:4) (69Koc) and its incomplete sub-unit (5:2) cluster, based on diffraction data at equilibrium at 900 °C (86Gar**1**) following a simulation of order parameters (clustering similar to P″).

The authors noted that the (10:4) blende ZnS cluster envisaged by Grimes *et al.* (84And) (86Grim) is not the most likely because it was characterized without taking into account the geometric relaxation in the crystal structure.

**Tomlinson and Catlow** (89Tom) attempted to relate the structural modeling of short-range order, and the macroscopic variation of $z$ as a function of p′, adjusted « by mass-action treatment ». The binding energy per vacancy (*See* (89Tom) Table 2 p. 66) leads they to order the clusters as follows: [(8:3)$^{+1}$, -2.24 eV], [(12:4), (6:2), (16:5)$^{-1}$, (12:5)$^{-1}$, (13:4)$^{-2}$, (4:1)$^{-1}$, (10:4)$^{-4}$, -2.22 to -1.68 eV], then [(5:2 M), -1.24 eV] and [(5:2 G), -0.71 eV]. Finally, they proposed a mixture of clusters [(12:4), (6:2), (4:1)] for the incommensurate «sub-phase P′».

**R. W. Grimes *et al.*** (84And) (99Min) considered the structure and stability of clusters with different coordination types. They base their analysis on a generalized crystal field for the d electrons of $Fe^{3+}$ and $Fe^{2+}$, including the energy of orbitals 2s and 2p of oxygen. Their computation of the binding energy per vacancy favors the clusters <110> ZnS blende and <111> spinel type stackings. It rules out clusters (8:3) (10:4) based on <100> stacking of (4:1) units, and the Koch and Cohen (13:4) cluster. Within the lower field of stoichiometry (P′ phase) smaller clusters coexist including (4:1), (6:2), (7:2) <110> and (10:4) ZnS blende types, (7:2) <111> spinel type. Larger clusters from the (12:4) <110> type, (18:6), (16:5) spinel <111> type, (16:5) <110> type and (16:7) <110> ZnS blende type clusters can be invoked at higher nonstoichiometry (P″ phase). The clusters « should be neutral in order to be in agreement with experimental data ».

NOTA: The classification of the clusters became more complicated because of the increasing number proposals in the literature. Initially, only the criterion of corner or edge sharing of the basic (4:1) units was needed. Mixed cases were thought to be the exception. Lebreton and Hobbs (83Leb) introduced three way of clustering: *edge sharing along <100> (Type I)*, *corner sharing along <110> (Type II)*, and *corner sharing along <111> (Type III)*. Mixed ways are also to be envisaged. Clusters based on composition alone are possible exceptions (*e.g.*, cluster (10:3)). Labidi and Monty (*See* (91Lab) p. 100-101) resumed the situation differently with the *types 1 (face shared) aligned in <100> direction, 2 (edge shared) aligned in <110> direction, and 3 (corner shared) aligned in <111> direction*. Worral and Coley (*See* (R-13Wor) p. 24-26) classified the clusters in three types defined as either *type 1 (corner shared as in magnetite)*, *type 2 (edge shared sharing one octahedral vacancy)*, or *type 3 (edge shared sharing two octahedral vacancies)*.

## III. Structural approaches. Models for short- and long-range ordering

### 1- Roth model: cluster (2:1)

With regards to experimental uncertainties inherent to X-ray diffraction on polycrystalline materials, it is necessary to use neutron diffraction experiments to determine correctly the site occupancies relative to iron atoms in octahedral and tetrahedral sites of the NaCl-like lattice. Initial results relative to quenched wüstite are obtained using neutron diffraction by W. L. Roth (60Rot) who determined the octahedral and tetrahedral site occupancies of vacancies and $Fe^{3+}$ cations, respectively represented by ($z+t$) and *t*. The characterization of the ratio R = ($z+t$)/$t$ was found close to R=2 for samples quenched from high temperatures. The neutron diffraction patterns obtained at 290 K and 4.2 K suggested the presence of defects constituted of two cation vacancies associated with one interstitial cation in the tetrahedral site. The average magnetic moment per cation site was found to be much smaller than expected.

### 2 - The Manenc's observations of phases P, P′, P″

Using X-ray diffraction analyses on quenched wüstites, a first structural determination of pseudo-phases was performed by Manenc *et al.* (63-68Man). This work is reviewed above in section II-2 §7. The most important result in this seminal work consisted in the description of three types of wüstite P, P′ and P″**,** observed in quenched samples.

### 3- The historical Koch and Cohen (13:4) cluster

In 1969, a study by X-ray diffraction was published by Koch and Cohen (69Koc) on a single crystal of Fe$_{0.902}$O ($z$=0.098) obtained after quenching from 1000 °C, and « *corresponding to the P′ phase of Manenc* ». The authors proposed a new commensurate defect structure based on (13:4) large clusters (Figure III-1) distributed in the fcc lattice (cell parameter a) with a repetition distance (3a, 3a, 3a) noted as 3X. For the first time, a detailed crystallographic determination proposed a cluster corresponding to the corner sharing <110> type agglomeration of four basic (4:1) clusters.

To perform their structural refinements, the authors fixed this cluster model assuming that the resulting superstructure should have a periodicity of 3X (clusters being arranged only with regular distances of three cell parameters), in agreement with the known composition $z$=0.098, despite the fact that the additional superstructure peaks correspond to a superstructure 2.6X incompatible with their assumption.

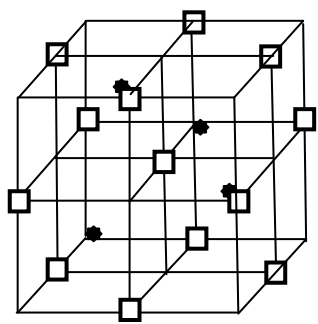

Fig. III-1 - The Koch and Cohen (13:4) cluster associated with a 3X commensurate structure of wüstite Fe$_{0.902}$O.

**4-** Using *in situ* neutron diffraction, **Cheetham et al.** (71Che) showed that the ratio R = ($z+t$)/$t$ was in the range between 3 and 4, with a mean value of 3.2, which is compatible with a specific (13:4) cluster (R = 3.25). Using neutron diffraction coupled with magnetic scattering, Battle and Cheetham (79Bat) determined ratios R = 2.78, 2.82, and 3.01 and concluded that a cluster (8:3) is consistent with their observations. They coupled neutron diffraction experiments to Mössbauer effect analyses (*See* (79Bat), Fig. 4 p. 343) and studied the antiferromagnetic coupling of iron cations in octahedral and tetrahedral sites of the FeO lattice. For the first time, these authors showed that atomic clustering can be characterized from the study of magnetic properties of clusters in a lattice. This result is used again in (13Sai).

**5-** The numerous results obtained by **Radler et al.** (90Rad) from neutron diffraction patterns of polycrystalline samples under equilibrium yield values of **R > 3** corresponding to mixtures of clusters at high temperature, and **R < 3** to clusters larger than (10:3) below 1173 K.

**6- Schweika et al.** (92- 95Sch) found a ratio of vacancies to interstitials R = 4.0 (±0.5) obtained under *in situ* conditions by neutron diffuse scattering from a single crystal with composition z = 0.08, at T=1150 °C. Their analysis of the diffuse scattering patterns leads them to conclude that 30% of the vacancies are free, while a further 15% of the defects are bound in isolated (4:1) defect clusters. In their analysis (92Sch), the authors stated that their results invalidate the drawn predominance of large clusters at elevated $z$. For example, the (13:4) cluster cannot fit the diffuse scattering that they characterized near the origin.

**7- Welberry and Christy** (95- 97- 14Wel) present a detailed study of X-ray diffuse scattering for a single crystal with composition $z$=0.057, obtained after quenching from a temperature which is not specified. Their sample corresponds either to the Manenc's P phase (S*ee* (68Man) and (77And)), or the W1 domain proposed by Vallet *et al.* (64- 79- 89Val). The X-ray analysis coupled with a « paracrystalline » modeling gives information on defect distribution, defect cluster size, number of interstitials, and lattice strain. The authors describe (statistically) a specific crystal constituted of ordered zones of clusters randomly distributed in a fcc lattice.

**8- Saines et al**. (13Sai) studied by neutron diffraction a polycrystalline sample of wüstite Fe$_{0.902}$O quenched from 900 °C (from the domain of W'). Their proposed nuclear and magnetic clusters are derived from an elaborate Reverse Monte Carlo modeling technique. The nuclear structure is described by islands of V4T units connected along the <110> directions « into a Koch- Cohen arrangement ».

The magnetic structure in the bulk between the clusters is described as a « non-collinear variant » of the antiferromagnetic structure along the [111] axis as previously envisaged (79Bat). Free vacancies are near the clusters. This structural approach can be compared with the paracrystalline description of the defect distribution by Welberry *et al.* (95Wel) (14Wel)**.**

## 9- Gavarri-Carel-Weigel's works

In 1979, using *in situ* neutron diffraction experiments, Gavarri *et al.* (79Gav) develop an analysis of the structural evolution in CO2/CO environments at high temperature. A systematic determination of the ratio of the vacancy rate $(z+t)$ divided by the interstitial rate $t$, R= $(z+t)/t$, is carried out. Quasi constant values of R close to 2.4 ± 0.4, at two equilibrium temperatures (985 and 1075 °C) and compositions $z$ ranging between 0.058 and 0.120, are determined, with standard errors induced by the uncertainties on the separation between Bragg peaks and a complex diffuse scattering.

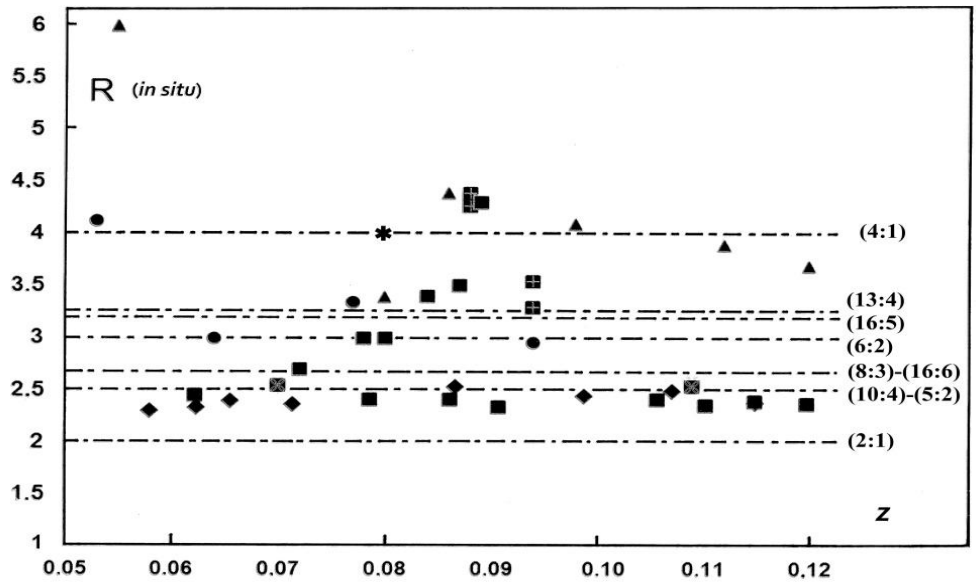

**Fig. III-2 – R =$(z+t)/t$ values *vs* $z$, determined under equilibrium, from the literature:**
●: at 800°C and ▥: at 925, 1050, 1125, 1180°C Cheetham *et al.* (71Che);
◆: at 1075°C and ■: at 985°C Gavarri *et al.* (79Gav); ▩: at 900°C Gartstein *et al*. (86Gar1)
▲: at 1050°C and ▪: at 850-1100°C Radler *et al.* (90Rad); ✱: at 1150°C Schweika *et al.* (95Sch).

The Figure III-2 displays the values of R obtained by different authors. In the case of studies made in situ, most of the values for R are between 2.4 and 3.5. The authors ((79Gav) Fig. III-3) observe two types of signals: (i) Bragg peaks from which the ratio R is determined, (ii) additional scattering corresponding to superstructure peaks, and thermal and static diffuse scattering associated with defect clustering, disorder and static distortions. The intensities $I_A(T,z)$ of this additional scattering increases linearly with composition $z$ and decreases with temperature. The ratios $I_A/I220$ where $I_{220}$ is the intensity of the (220) Bragg peak vary as follows:

- $I_A/I220$ (985 °C) = 0.19 + 4.05($z$-0.062)
- $I_A/I220$ (1075 °C) = 0.15 + 4.25 ($z$-0.058)

Therefore, these linear variations coupled with a small variation of the superstructure parameter k suggest a quasi-invariance of cluster size with clusters agglomerated in domains, and with the size of these domains increasing with $z$. The position of the centroid of this scattering allows the determination of an average superstructure corresponding to an irregular spacing of clusters quasi-constant with $z$ or T. The average distance between clusters is evaluated to $2.7a_0$, $a_0$ being the cell parameter of ideal wüstite for a fixed temperature.

In later work on quenched samples of magnesio- and calciowüstites by neutron diffraction, using Rietveld refinements, Gavarri *et al.* (87Gav) (90Car) determine R values close to 2.5 for pure wüstites or for magnesiowüstites $Fe_{(1-z-y)}Mg_yO$. In the case of calcio-wüstites $Fe_{(1-z-y)}Ca_yO$, as the composition $y$ increases, the ratio R increases strongly with R > 4 and the superstructure vanished. Under these conditions, the cluster sizes is strongly modified, and a majority of free vacancies likely appears. These studies demonstrate that the formation of clusters is stabilized by Mg doping, and destabilized by Ca doping, producing free vacancies that coexist with residual clusters.

It should be noted that the diffuse scattering due to long-range ordering is similar in diffraction patterns of quenched and *in situ* samples. Consequently, only a slight difference exists between the nature of defects and their organization in the fcc lattice between high temperatures and room temperature.

**Table III-1: R = $(z+t)/t$ values from quenched substituted wüstites (90Car).**

| $Fe_{1-z-y}$ $Mg_yO$ | Fraction $y$ (Mg) | R = $(z+t)/t$ |
|---|---|---|
| $z$=0.065 | 0.075 | 2.53(0.5) |
| $z$=0.085 | 0.030 | 2.88(0.6) |
| | | |
| $Fe_{1-z-y}$ $Ca_yO$ | Fraction $y$ (Ca) | R = $(z+t)/t$ |
| $z$=0.0688 | 0.010 | 2.20(0.5) |
| $z$=0.0649 (*) | 0.030 (*) | 3.31(0.5) |
| $z$= 0.0534 | 0.050 | 8.22(4.0) |
| $z$=0.085 | 0.010 | 2.37(0.5) |
| $z$=0.0899 | 0.030 | 3.39(0.5) |

(*) See Figure III-3

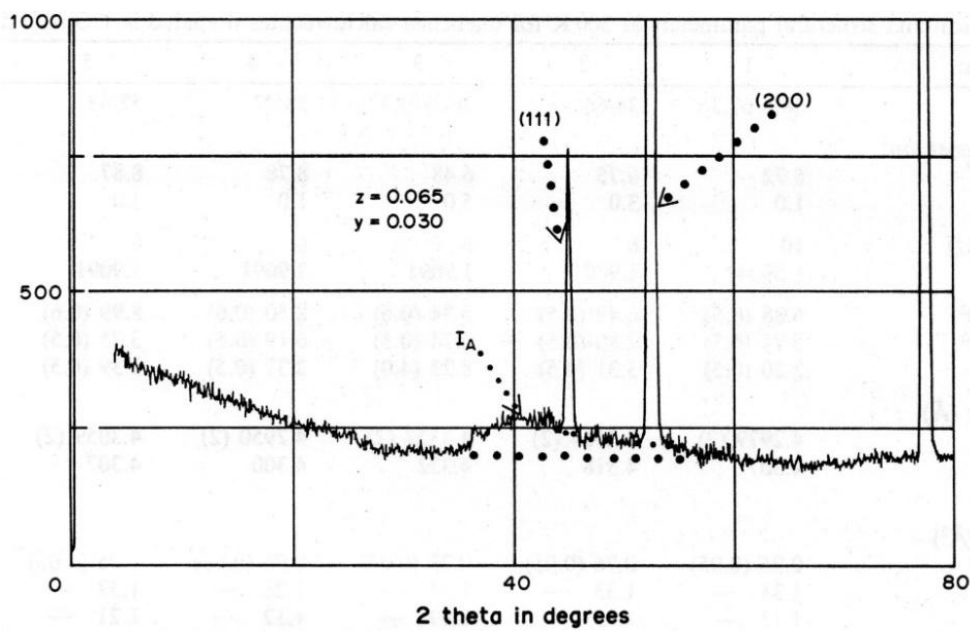

**Carel and Gavarri (90Car)**

**Figure III-3 - Neutron diffraction pattern ($\lambda$ = 1.9 Å) of a quenched polycrystalline calcio-wustite $Fe_{1-z-y}Ca_yO$ ($z$ = 0.065, $y$ = 0.03) in room conditions. Presence of a large diffuse scattering ($I_A$) due to the superstructure (k close to 2.7), and (110) small peak (36° 2θ) due to vacancies and interstitial $Fe^{3+}$ in the lattice. The intensity of $I_A$ decreases with the calcium fraction $y$. Cell parameter a(25 °C) = 4.3164 Å. R = $(z+t)/t$ = 3.3.**

Table III-1 above lists the values of R corresponding to the quenched pure and substituted wüstites resulting from these studies.

Figure III-3 shows an example of a neutron diffraction pattern obtained from a quenched polycrystalline calciowüstite ($z = 0.065$; $y = 0.030$). In addition to Bragg peaks, the complex signal due to the superstructure and the structural disorder can be observed. For this sample, $R = 3.3$ ($\pm 0.5$).

The values of the ratio $R = \mathbf{(z+t)/t}$ obtained from the literature on quenched pure and Ca or Mg substituted wüstites are displayed in Fig. III-4. The utmost value $R = 8.2$ was obtained for a calcium-rich calciowüstite ($y=0.05$).

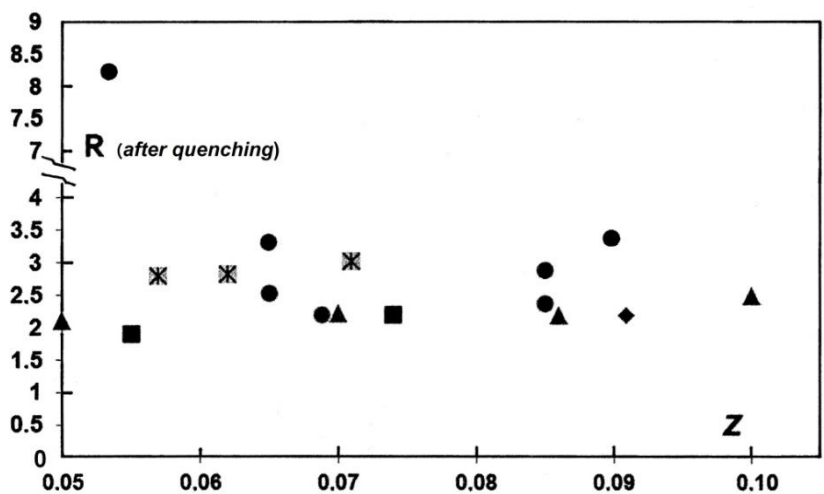

Fig. III-4 - R values *vs* z from literature determined after quenching.
WMg and WCa, ● = Carel and Gavarri (90Car); pure Fe1-zO: ■ = Roth (60Rot);
▲ = Smuts (66Smu); ◆ = Gavarri *et al*. (79Gav); ✱ = Battle and Cheetham (79Bat).

It is to be noted that **Gleitzer and Goodenough** suggested that the ratio R=2.4 ± 0.4 found by Gavarri *et al*. for all values of *x* «... *seems to be indicative of a single structural principle. This result does not establish the structural binding block uniquely, but it does seem to eliminate several alternatives such as the Koch Cohen cluster and the isolated 4:1 unit* » (*See* (R-85Gle) p. 51-54).

**10- Transmission Electron Microscopy analyses**

In the past, microstructural analyses using electron diffraction and HREM images were performed. See in particular the analyses of Andersson and Sletnes (77And) (*See* II 4: p. 18). The images obtained by Iijima (74Iij) for a sample Fe$_{0.92}$O (*provided by Prof. Cohen*) are often referenced.

**11- The hypothetical cluster (10:4) <110> type with a ZnS-blende configuration**

When re-interpreting the images obtained from electron microscopy experiments by Ishiguro and Nagakura (85Ish), Gavarri *et al*. (88Gav) then Nihoul *et al*. (91Nih), show that the possible cluster (10:4) corresponding to the ideal ratio $R = 2.5$ is compatible with all the experimental data available (at least for quenched wüstites). This cluster is formed of a « FeO » cubic cell surrounded by an envelope of $Fe^{3+}$ cations with total 3D dimension (2a x 2a x 2a).

Figure III-5 depicts this cluster. The authors (88Gav) and (91Nih) show that image simulations from this cluster model are in good agreement with the high-resolution images obtained by Ishiguro and Nagakura (85Ish), in the case of crystals quenched from 1000 °C, and with a composition $z = 0.10$.

The formation of local ordered domains of these clusters can be at the origin of all observations corresponding to a mean ratio R close to 2.5, and a quasi-invariant superstructure. The authors propose a progressive ordering of clusters as *z* increased: for low *z* values, repetition distances 3a linkable to the P variety by Manenc, or W1, are observed. For intermediate $z$ values, repetition distances 3a and 2.55a (monoclinic associations) linkable to P′ or W2 appear. Finally, ordered superstructures (2.55a x 2.55a x 2.55a) are formed in local zones linkable to W3 or P″. Specific well-ordered superstructures can be based on a 3D super-structure with cell parameters (5a x 5a x 5a). This superstructure can be associated with the limit evolution at the boundary W3/Fe3O4. Such a superstructure was depicted by Shaïovitch and Zvinchuk (87Sha). Finally, the distinction between « varieties of wüstite » or pseudo-phases can be due to both modifications of long-range order of clusters and the coexistence of different clusters that include free vacancies.

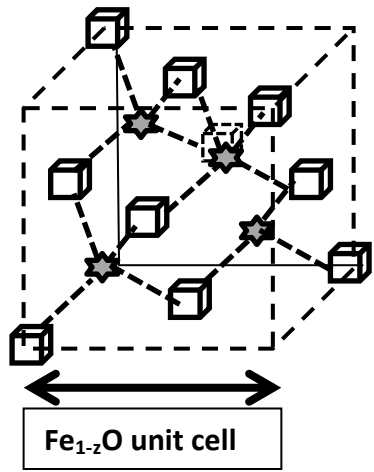

Fe$_{1-z}$O unit cell

**Fig. III-5 - The cluster (10:4) ZnS blende <110> type.**

✪ = $Fe^{3+}$ in tetrahedral site

▱ = Iron vacancy in octahedral site

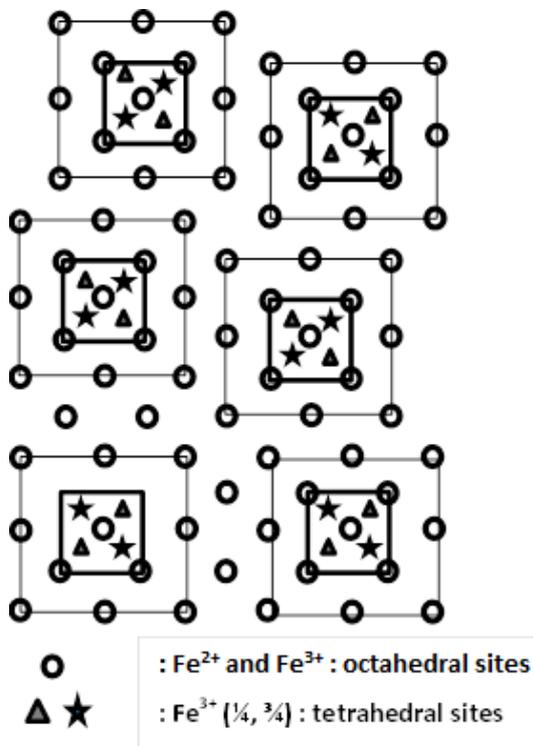

○ : $Fe^{2+}$ and $Fe^{3+}$ : octahedral sites

▲ ★ : $Fe^{3+}$ (¼, ¾) : tetrahedral sites

**Fig. III-6 - Various orderings of (10:4) clusters in the fcc lattice of wüstite. The basic cluster can be the (13:4), (10:4) or other possibilities with 4 interstitial sites occupied by cations $Fe^{3+}$. Disorder can develop by statistical displacement in one, two or three directions. 2.55a could be the smaller inter-cluster distance (a = cell parameter of the "FeO" lattice).**

To define an isolated cluster correctly, it is necessary to take into account its envelope of $Fe^{2+}$ and $Fe^{3+}$ cations in octahedral sites. The charge of the cluster can be postulated as a first approximation by assuming the individual charges of vacancies and of iron atoms. In Figure III-6 (on the left), we illustrate the possible arrangement of these ZnS blende defect cluster (10:4), in the fcc lattice. The minimum distance between two clusters is $[(2.5a)^2 + (0.5a)^2]^{1/2} = 2.55a$, and with allowed distances up to 3a. This model is based on a cluster core (10:4) with double envelope of octahedral iron cations. If we assume that a cluster (m:n) has a global charge Q= 3n-mq (m vacancies with formal charges q (=2,1,0) and n interstitials $Fe_i^{\circ\circ\circ}$ with charges +3), then the (10:4) cluster has a charge of Q= -8, +2, or +12 for q= 2, 1 or 0, respectively. The electroneutrality is ensured by envelopes with charges -Q. Using this type of defect structure, a model of ordered (10:4) clusters was proposed (88Gav2 and 91Nih), thus allowing simulation of HREM images in the two directions [001] and [101] of the crystal. These hypothetical distributions of clusters (Fig. III-7) are in good agreement with the HREM images by Ishiguro and Nagakura (85Ish).

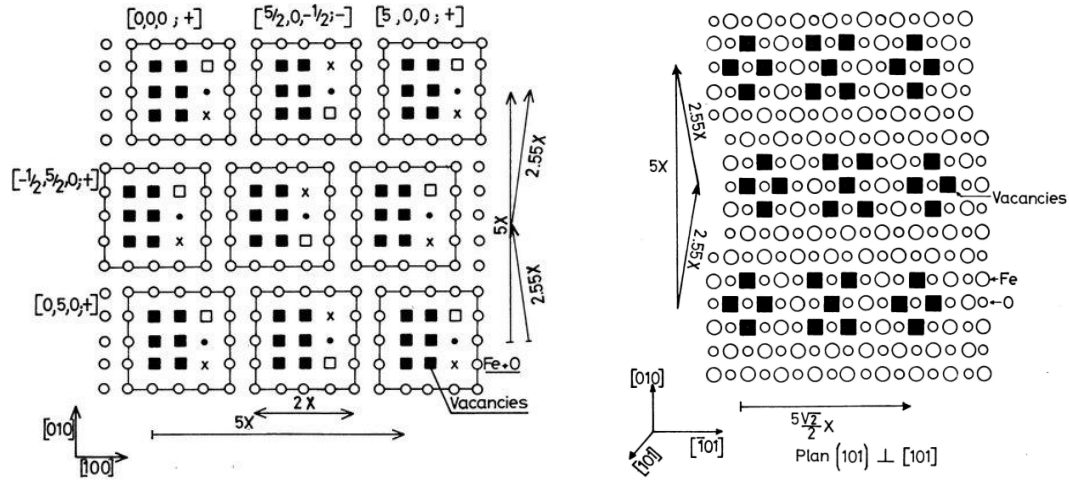

*by* Gavarri *et al.* (88Gav)

**Fig. III-7 – Model of local ordering of (10:4) blende ZnS clusters by (88Gav) (91Nih). It allows reproducing the experimental HREM images from (85Ish).**

### IV- Hypothesis of a mixture of clusters and/or of cluster zones

#### 1- Mixture of defect clusters

Based on previous publications by Gavarri and Carel (79- 81Gav), a new specific relationship can be established taking into account ordered zones of clusters and zones in which free vacancies are distributed. A mixture of different clusters $(m:n)_1$ and $(m:n)_2$ can coexist with additional free vacancies. As a first approximation, the authors consider a lattice based on N cells of $Fe_{1-z}O$, with a proportion $V(z)$ of zones without defect. If $N_i$ is the number of cells where clusters $(m:n)_i$ are formed, with $m_i$ and $n_i$ the characteristics of a specific cluster, and $k_i.a$ the distance between two clusters, then the relation between cluster sizes and long-range order in the case of the fcc lattice of wüstite can be expressed as follows:

$$1-V(z) = 4z \cdot \Sigma_i \{ \pi_i (k_i)^3 / (m_i-n_i) \} \text{ with } \pi_i = N_i/N$$
(IV-1)

For a unique cluster with continuous order (high $z$ values), the relation (IV-1) yields:

$$(m-n) / (k)^3 = 4z \quad \rightarrow \quad (m-n) = 4zk^3$$
(IV-2)

Considering the experimental values of k available in the literature, ranging between 2.5 and 3, this parameter k is characteristic of zones in which clusters are ordered partially: this means that, for low values of $z$, zones without clusters exist and that they can be occupied by isolated defects (*i.e.*, free vacancies).

In this case, an alternative simplified approach can be proposed. If we assume that ordered zones are formed progressively in the lattice as $z$ increases, and that free vacancies (with local fraction $z_{free}$) coexist with and outside of these cluster zones, then we can express a simple relation between the value of z and the effective fraction of vacancies in cluster zones $z_{Cluster}$ using the repartition coefficients $\alpha_1$ for cluster zones and $\alpha_2$ for free vacancies zones.

$$z = \alpha_1 \cdot z_{Cluster} + \alpha_2 \cdot z_{free} \quad \text{with} \quad \alpha_1 + \alpha_2 = 1$$
(IV-3)

For a fixed equilibrium temperature, the phase diagram allows the determination of the compositions $z_{min}$ and $z_{max}$ corresponding to the existence of wüstite. It can be shown easily that the coefficients $\alpha_1$ and $\alpha_2$ are expressed as follows:

$$\alpha_1(T) = (z - z_{min})/(z_{Cluster} - z_{min})$$
$$\alpha_2(T) = 1 - \alpha_1(T)$$
(IV-4)

The coefficient $\alpha_1(T)$ is equal to zero for $z = z_{min}$ (no cluster zone) and equal to $(z_{max} - z_{min})/(z_{Cluster} - z_{min})$ for $z = z_{max}$. The concentration $z_{Cluster}$ can be approximated as $z_{Cluster} = z \cdot (k_1)^3/(m_1 - n_1)$.

This simplified approach can explain a large part of the experimental observations in the literature. For low $z$ values, only a few cluster zones are formed, coexisting with free vacancies distributed between these zones. The major part of the crystal lattice is dominated by disordered point defects, and electrical conduction is due to holes.

For intermediate values of $z$, the cluster zones join together, forming percolation paths of ordered zones. A mixed conduction mode is expected. For high values of $z$, the lattice is occupied fully by ordered clusters, with likely small changes in the arrangement of clusters (variable k values from 3 to 2.5). The conduction is conditioned by neutral vacancies and electrons in conduction band.

Using this model, we can recognize clearly the W1 disordered phase with p type conduction, the intermediate phase W2 as a mixture of cluster zones and free vacancy zones with mixed conduction, *i.e.*, (holes +) and (electrons -) coexisting at short distance, and W3 as an ordered phase with conduction of n type from electrons in the conduction band.

## 2- Variation of parameter k

Numerous structural data can be found in the studies by **Bauer** *et al.* Part I (R-80Bau) and Part II (80Bau). These authors develop a series of data set using X-ray diffraction on powders of quenched wüstites. They observe linear variations of the cell parameter as a function of composition $z$, and of the position of superstructure peaks. They characterize these peaks with parameter $\delta = 1/k$, where k designates the superstructure parameter discussed previously. Parameter $\delta$ varies between 0.37 and 0.39, corresponding to values of k between 2.7 and 2.56. This variation observed in quenched samples is in good agreement with the model presented in section IV-1 with a minimum k value of 2.55. The authors (80Bau) considered that their data did not permit to observe any phase transition. However, they suggested that wüstite could be described by an incommensurate lattice of clusters with the presence of different clusters based on cluster (4:1).

## 3 - Statistical analysis of diffuse scattering and new cluster

**Gartstein and Mason** (82Gar) showed that p conduction was observed in wüstite and that electrical conduction was based on a hopping mechanism between adjacent clusters. Later, Gartstein *et al.* published a crystallographic analysis of a $Fe_{1-z}O$ ($z=0.07$) single crystal using X-ray diffraction at 1173 K (86Gar**1,2**). They developed a detailed analysis of diffuse scattering and Bragg peaks, and concluded to the existence of a mixture of clusters (13:4) and asymmetric new clusters (5:2) constituted of 5 vacancies and 2 interstitials corresponding to the ideal ratio R = 2.5. Their data are in agreement with the values found by (79Gav). Gartstein *et al.* proposed an ordering of these clusters with an average repetition distance of 2.5a, close to the hypothetic value of 2.55a proposed earlier (88Gav**)** (91Nih).

**4- Summary.** As a conclusion to this section, it should be noted that there is an inherent difference between the concept of a mixture of disordered clusters with various forms, and the concept of zones with identical disordered clusters, coexisting with zones of free point defects. The concept of incommensurability, applied to specific compositions $z$, was also proposed as a mathematical approach of disorder (82Yam) (87Wei). Finally, the proposition of the irregular cluster (5:2), resulting from the analysis of diffuse scattering by (86Gar**1,2**), seemed to confirm the quasi-constant value $R = (z+t)/t = 2.4 \pm 0.4$ proposed by us.

# V - Defect clustering and equilibrium equations

In this section, the well-known classical models of defect equilibria in the case of cation deficient non-stoichiometric oxide $Fe_{1-z}O$ are reviewed. In addition, we propose new specific possible models applicable to wüstite (*See particularly* (R-81Sor) (85Sor)). A series of equilibrium equations of increasing complexity are presented, in relation with the historical evolution of the concepts concerning point defects in oxides.

In nonstoichiometric iron monoxide $Fe_{1-z}O$, where only $Fe^{2+}$ and $Fe^{3+}$ cations (also noted $Fe_{Fe}$ and $Fe_{Fe}°$) can coexist, cation vacancies can have one of three possible stable charges: doubly charged vacancies $V_{Fe}''$, simply charged vacancies $V_{Fe}'$, and neutral vacancies $V_{Fe}^X$ (noted $V_{Fe}$). Singly charged iron vacancies $V_{Fe}'$ were proposed previously in place of doubly charged vacancies $V_{Fe}''$ by Kofstad and Hed (68Kof) (*See* II §15). Oxygen vacancies are not taken into account because they occur in negligible numbers compared to the iron vacancies.

Three hypothetical equilibria previously presented in section II (in which the activity of oxygen $O_O$ in the solid is taken equal to 1) are listed below, with q=0, 1 or 2 and p′ = p(O2)

$$½ O_2 \leftrightarrow O_O + V_M^{q(')} + q\ h° \quad (V-1)$$

$$K_q = [V_{Fe}^{q(')}][h°]^q / p'^{½}$$

Considering the relationship $-q.[V_M^{q(')}]+[h°] = 0$ (electroneutrality condition) with $[V_{Fe}^{q(')}] = z$, we can derive a general expression in which s =2(q+1):

➡ $\log z = 1/s . l' . \log (K/C)^2$ (with $C = (q)^{2q}$)

The three hypothetical values of s corresponding to q = 2, 1, and 0 are s = 6, 4, and 2, respectively.
The modifications of vacancies can be described in terms of two equilibriums involving holes h° in the valence band

$V_{Fe}'' + h° \leftrightarrow V_{Fe}'$      (V-1a)

and    $V_{Fe}' + h° \leftrightarrow V_{Fe}$      (V-1b)

Therefore, three domains corresponding to the three types of iron vacancies exist in all systems presenting significant cation deficiency.

In the case of $Fe_{1-z}O$, additional reactions occur since two valences ($Fe^{2+}$ and $Fe^{3+}$) exist:

$Fe_{Fe} + h° \leftrightarrow Fe_{Fe}°$ (*$Fe^{3+}$ in octahedral sites*)

and    $Fe_{Fe}° \rightarrow Fe_i°°°$ (*$Fe^{3+}$ in tetrahedral sites*)
                 + $V_{Fe}''$ (*additional cation vacancy*)

As a first approximation, assuming that the sole point defects are iron vacancies associated with $Fe^{3+}$ ($Fe_{Fe}°$) and coexisting with $Fe^{2+}$ ($Fe_{Fe}$) in $Fe_{1-z}O$, the general relationship between composition $z$ and oxygen partial pressure p′ is as follows

$q\ Fe_{Fe} + ½ O_2 \leftrightarrow q\ Fe_{Fe}° + V^{q(')} + O_O$   (V-2)

with the equilibrium constant:

$K_q = [Fe_{Fe}°]^q . [V^{q(')}] / ([Fe_{Fe}]^q . p'^{½})$

In this expression $[Fe_{Fe}] = 1-3z$ and $[Fe_{Fe}°] = 2z$. The activity of $O_O$ is taken equal to 1.

As all molar fractions are related to the composition $z$, we can derive the relationship:

$K_{IV-2} = [Fe_{Fe}°]^q . [V^{q(')}] / ([Fe_{Fe}]^q . p'^{½})$
        $= 2^q . z^{(q+1)}/(1-3z)^q . p'^{½}$

If this simple model is correct then the term $(1-3z)$ should vary slowly with $z$ in a reduced composition range, meaning that the activity of $Fe_{Fe}$ can be considered to be constant. Under this assumption, $z$ is roughly proportional to $(p')^{1/s}$, with s = 2(q+1) being an integer characteristic of the nature of the defects. For q = 2 (doubly charged vacancies) we obtain s ≈ 6. For q = 1 (singly charged vacancies), s ≈ 4. For neutral vacancies, s ≈ 2.

If electrons are present in the system, and keeping the same approximations as above:

$2\ Fe_{Fe} + ½ O_2 \leftrightarrow 2\ Fe_{Fe}° + V^{q(')} + (2-q)\ e' + O_O$
                           (V-3)

the equilibrium constant is

$K_{IV-3}(q) = [Fe_{Fe}°]^2 . [V^{q(')}] . [e']^{(2-q)} / ([Fe_{Fe}]^2 . p'^{½})$

For q=0 (neutral V), $z$ is roughly proportional to $(p')^{1/10}$, and s = 10; for q=1 (simply charged vacancies), s = 8; for q=2, s = 6.

In the case of cations $Fe^{3+}$ located partially in tetrahedral sites, and using the chemical formula

$$[Fe_{Fe}]_{1-3z} [Fe_{Fe}°]_{2z-t} [Fe_i^{°°°}]_t [V_{Fe}'']_{z+t} O_O,$$

equation (V-3) becomes:

$$2 Fe_{Fe} + \tfrac{1}{2} O_2 \longleftrightarrow Fe_{Fe}° + Fe_i^{°°°} + 2V_{Fe}'' + O_O \quad (V\text{-}4)$$

giving rise to the equilibrium constant (with all appropriate approximations):

$$K_{IV\text{-}4} = [Fe_{Fe}°][Fe_i^{°°°}][V_{Fe}'']^2 / p'^{1/2} = (2z-t)(t)(z+t)^2 / ([Fe_{Fe}]^2 . p'^{1/2})$$

Based on X-ray and neutron diffraction experiments from various authors (*See* section III), the ratio R = *(z+t)/t* was found to be between 2 and 4. Consequently, if we assume that $t = z/(R-1)$ is roughly proportional to $z$ (*i.e.* R is quasi constant), and term $[Fe_{Fe}]^2$ varies slowly, then $z$ is roughly proportional to $p'^{1/8}$ yielding a value s = 8.

If we take into account the existence of free electrons associated with the modification of vacancy charges (noted q(·) or -q, we can express the additional equilibrium:

$$2 Fe_{Fe} + \tfrac{1}{2} O_2 \longleftrightarrow Fe_{Fe}° + Fe_i^{°°°} + 2V^{q(/)} + 2(2-q))e' + O_O \quad (V\text{-}5)$$

$$K_{IV\text{-}5} = [Fe_{Fe}°][Fe_i^{°°°}][V_{Fe}^{q(/)}]^2[e']^{2(2-q)} / ([Fe_{Fe}]^2 . p'^{1/2})$$

For q = 2, 1, 0 → s = 8, 12, 16

The notion of clusters of defects can be added to this model using a simplified description to avoid excessive complications. In the case of clusters sharing envelopes (for high values of $z$) the notion of isolated clusters is not strictly valid. As a first approximation in this case, the charge of a cluster (noted Q, either positive or negative) is calculated from the individual charges of $Fe^{3+}$ in tetrahedral sites and from octahedral iron vacancies. Under these hypotheses, we can express the formation of simplified clusters as:

$$2 Fe_{Fe} + \tfrac{1}{2} O_2 \longleftrightarrow Fe_{Fe}° + \{Fe_i^{°°°}, V_{Fe}^{q(/)}\}^Q + V_{Fe}^{q(/)} + 2(2-q))e' + O_O \quad (V\text{-}6)$$

$$Q = 3 - q$$
$$K_{IV\text{-}6} = [Fe_{Fe}°].[\{Fe_i^{°°°}, V_{Fe}^{q(/)}\}].[V_{Fe}^{q(/)}].[e']^{2(2-q)} / ([Fe_{Fe}]^2 . p'^{1/2})$$

In this equilibrium, we introduce simplified clusters $\{Fe_i^{°°°}, V_{Fe}^{q(/)Q}\}$ with hypothetical electrical charges Q=3-q (q=0,1,2), coexisting with free $Fe^{3+}$ (or $Fe_{Fe}°$) in octahedral sites. As $z$ increases, it is reasonable to consider that all fractions of species $[Fe_{Fe}°]$, $[Fe_i^{°°°}, V_{Fe}^{q(/)}]$, $[V_{Fe}^{q(/)}]$, and [e'] are proportional to $z$ and that the equilibrium constant can be expressed using a term C depending on activity coefficients as follows:

$$K_{IV\text{-}6} = C [z^3 . z^{2(2-q)}] / [(1-3z)^2 . p'^{1/2}]$$

If we consider composition domains in which the activity coefficients of $Fe_{Fe}$ vary slowly as a function of $z$, term C can be considered to be constant. For intermediate and high values of $z$, the function $z^{(7-2q)} / (1-3z)^2$ can be approximated by an average function $z^X$ with 3<X<4 for q=2, 4<X<5 for q=1, and 6<X<7 for q=0. Consequently, the values of s are in the range 6< s <8 (for low $z$ values and q=2), 9< s <10 (for intermediate $z$ values and q=1) and 11< s <12 (for high $z$ values q=0).

As a supplementary step, a cluster can be formed from point defects $Fe_{Fe}°$, $Fe_i^{°°°}$, $V_{Fe}^{q(/)}$ with $Fe_{Fe}°$ in the envelope of the basic cluster, yielding:

$$2 Fe_{Fe} + \tfrac{1}{2} O_2 \longleftrightarrow \{Fe_{Fe}°, Fe_i^{°°°}, V_{Fe}^{q(/)}\}^Q + V_{Fe}^{q(/)} + 2(2-q))e' + O_O \quad (V\text{-}7)$$

$$Q = 4-q$$
$$K_{IV\text{-}7} = [\{Fe_{Fe}°, Fe_i^{°°°}, V_{Fe}^{q(/)}\}^Q][V_{Fe}^{q(/)}][e']^{2(2-q)} / [Fe_{Fe}]^2 . p'^{1/2}$$

In this equilibrium, we consider clusters $\{Fe_{Fe}°, Fe_i^{°°°}, V_{Fe}^{q(/)}\}$ with their $Fe^{3+}$ envelopes. As $z$ increases, it is reasonable to consider that all fractions of species $[\{Fe_{Fe}°, Fe_i^{°°°}, V_{Fe}^{q(/)}\}]$, $[V_{Fe}^{q(/)}]$ and [e'] are proportional to $z$ and that the equilibrium constant can be expressed with a specific term C depending on activity coefficients as follows:

$$K_{IV-7} = C [z^2 \cdot z^{\,2(2-q)}] / [(1-3z)^2 \cdot p'^{1/2}]$$

If we consider composition domains in which the activity coefficients vary slowly as a function of $z$, the term C can be considered to be constant. For intermediate and high values of $z$, the function $z^{(6-2q)}/(1-3z)^2$ can be approximated by the function $z^X$ with $X < 2$, 4 or 6, respectively, for q=2, 1, or 0 (See eq. (IV.6) above). Therefore, the values of s will be lower than (but close to) s=4 (for low values of $z$ and q=2), s=8 (for intermediate values of $z$ and q=1) and s=12 (for high values of $z$ and q=0).

**Table V-1a,b: Characteristic exponents s associated with different elemental models of defect equilibria, expressed in the form: $\log z \approx (1/s) \cdot l'$**

| 1a : Defects and hypothetical equilibrium equations of formation | |
|---|---|
| Eq. (V-2) | q Fe$_{Fe}$ + ½ O$_2$ ⇌ q Fe$_{Fe}^\circ$ + V$_{Fe}^{q(\prime)}$ + O$_O$ |
| Eq. (V-3) | 2 Fe$_{Fe}$ + ½ O$_2$ ⇌ 2 Fe$_{Fe}^\circ$ + V$_{Fe}^{q(\prime)}$ + (2-q) e' + O$_O$ |
| Eq. (V-5) | 2 Fe$_{Fe}$ + ½ O$_2$ ⇌ Fe$_{Fe}^\circ$ + Fe$_i^{\circ\circ\circ}$ + 2 V$_{Fe}^{q(\prime)}$ + 2(2-q))e' + O$_O$ |
| Eq. (V-6) | 2 Fe$_{Fe}$ + ½ O$_2$ ⇌ Fe$_{Fe}^\circ$ + { Fe$_i^{\circ\circ\circ}$, V$_{Fe}^{q(\prime)}$ } + V$_{Fe}^{q(\prime)}$ + 2(2-q))e' + O$_O$ |
| Eq. (V-7) | 2 Fe$_{Fe}$ + ½ O$_2$ ⇌ { Fe$_{Fe}^\circ$, Fe$_i^{\circ\circ\circ}$, V$_{Fe}^{q(\prime)}$ } + V$_{Fe}^{q(\prime)}$ + 2(2-q))e' + O$_O$ |

| 1.b: Calculations of s according to log z ≈ (1/s).$l'$ The fractions of defects are assumed to be proportional to $z$ in Fe$_{1-z}$O | | | | | |
|---|---|---|---|---|---|
| Charges q V$^{q-}$ | Exponent s Eq. (V-2) | Exponent s Eq. (V-3) | Exponent s Eq. (V-5) | Exponent s Eq. (V-6) | Exponent s Eq. (V-7) |
| 2 | 6 | 6 | 8 | 6 | 4 |
| 1 | 4 | 8 | 12 | 10 | 8 |
| 0 | 2 | 10 | 16 | 14 | 12 |
| Experimental agreement | Bad | Good | Bad | Unrealistic | Acceptable |

Given the preceding discussion, we note that, in the case of the example given in section II, these calculated values for s are close to the values observed in the literature. Table V-1a,b lists the hypothetical equations (1a) and the possible values of coefficient s characteristic of various possible equilibria following equation (1b).

**Complex clusters**

In this section, we introduce complex clusters (m:n)$^Q$ with variable formal charge Q and we consider successive reactions in which vacant tetrahedral sites V$_T$ are introduced:

2Fe$_{Fe}$+½O$_2$ → 2Fe$_{Fe}^\circ$ + V$_{Fe}^{q(\prime)}$ + (2-q)e'+O$_O$

(V-8a)

Fe$_{Fe}^\circ$ + V$_T$ → Fe$_i^{\circ\circ\circ}$ + V$_{Fe}^{q(\prime)}$ + (2-q)e'

(V-8b)

The formation of stable cluster :

nFe$_i^{\circ\circ\circ}$ + mV$_{Fe}^{q(\prime)}$ → { n Fe$_i^{\circ\circ\circ}$, m V$_{Fe}^{q(\prime)}$ }$^Q$

(V-8c)

with Q = 3n-qm (positive or negative charge).

The formal charge Q of a cluster is assumed to result from individual charges of vacancies with the invariant charges of interstitials Fe$^{3+}$ (Fe$_i^{\circ\circ\circ}$). As a first approximation, we can fix the charge of interstitial to simplify the calculations. However, the Fe$^{3+}$ interstitial can be modified into interstitials Fe$^{2+}$ interstitial cation due to electron jumping, and as a result a more complete model should take into account various possible Q values of cluster charges due to the modification of interstitial charges.

Using a more general hypothesis implying a cluster (m:n), and adding the presence of sites $V_T$ (tetrahedral vacant site), a more elaborate equation could be expressed as follows :

$2(M)Fe_{Fe} + nV_T + \frac{1}{2}(M)O_2 \leftrightarrow$
$\{nFe_i^{\circ\circ\circ}, mV_{Fe}^{q(\prime)}\}^Q + (2M-n).Fe_{Fe}^{\circ}$
$+ (M-m+n).V_{Fe}^{q(\prime)} + (M+n)(2-q).e' + (M).O_O$
(V-9)

with the hypothesis: $Q = 3n-qm$ (positive or negative).

In this equation, the integer $M \geq m$ is introduced arbitrarily to take into account the possible presence of additional free species $Fe_{Fe}^{\circ}$ and $V_{Fe}^{q(\prime)}$, associated with free electrons in the conduction band.

We have considered the tetrahedral sites as being initially the vacant sites $V_T$ susceptible to receive an interstitial $Fe_i^{\circ}$ cation.

As a major fraction of these sites (8 sites for 4 FeO units in a fcc cell) is vacant, their activity can be considered as being quasi-constant, and does not play a significant role in the calculation of equilibrium constants. The resulting equilibrium constant is as follows, with the reduced notation

$[CL] = \{n\ Fe_i^{\circ\circ\circ}, m\ V_{Fe}^{q(\prime)}\}^Q$
$K(T) =$
$[CL].[Fe_{Fe}^{\circ}]^{(2M-n)}.[V_{Fe}^{q(\prime)}]^{(M-m+n)}.[e']^{(M+n)(2-q)}$
$/ ([Fe_{Fe}]^{2M}.[V_T]^n.(p')^{M/2})$

Applying the same approximation as in previous equilibrium equations, the equilibrium constant can be expressed as follows with the additional hypothesis M=m:

$K(T) = z^{(2M-n)}.z^n.z^{(M+n)(2-q)}/((1-3z)^{2M}.(p')^{M/2})$
$\approx z^{(2m+1+(2-q)(m+n))}/(1-3z)^{2m}.(p')^{M/2}$

No activity coefficient is taken into consideration. Consequently and in a restricted range of composition $z$, no significant change of interactions occurs between point defects or clusters.

To illustrate the possible values of parameter s corresponding to the three hypothetical values for q, we have simulated $l'$ as a function of $\log z$ using the expressions of K(T) developed above, and determine the slopes of the curves ($l'$ vs $\log z$). We have chosen two specific clusters to calculate the various coefficients s depending on the charges q of vacancies. It result that:

- for a cluster (4:1), the three coefficients s are equal to 5.7, 8.2 and 10.7 for q=2, 1 and 0 respectively;

- for a cluster (10:4), the three coefficients s are equal to 5.4, 8.2 and 11 for q=2,1 and 0 respectively.

It should be noted that, for either cluster (4:1) or (10:4), coefficients s resulting from the calculations are quite similar for a given value of q. Given the approximations included in the model and expected experimental uncertainties, these simulated values (from s = 5 to 11) are congruent with the experimental s values obtained in the literature and ranging between 4 and 10. In other terms, the existence of pseudo-phases can be reasonably thought to be associated with the variation of the electrical charges (q-) of vacancies, and more generally with the variations of the electrical charge Q of the clusters. These changes of electrical charges are strongly correlated with the increasing density of them. Two neighbouring vacancies V″ with the same negative charge can produce two singly charged V′ or neutral V vacancies, which decreases the coulomb repulsion between them, and requires the addition of free electrons to the conduction band.

Now, using all the values of s discussed previously and resulting from diverse models, it is possible to justify the various modifications observed by many authors including Vallet *et al.* (64-79-89Val), Geiger *et al.* (66Gei), Fender and Riley (69Fen), Takayama and Kimizuka (80Tak), Worral and Coley (10- R-13Wor). In particular, let us recall the values s(W1) = 4.6, s(W2) = 6.2, s(W3) = 9.2, extracted from the data by Vallet et al., as well as values s(W1) = 4.8, s(W2) = 6, s(W3) = 9 extracted from the data by Takayama and Kimizuka (80Tak). Several equilibria might be in agreement with these experimental values of s. Among the possible solutions shown in Table IV-1, it is reasonable to consider the equilibria (V-3) and (V-7) in relatively good agreement with experiments, with q values varying from 2 to 0.

However, the general equation (V-9) based on clusters (m:n) seems to be in better agreement with thermodynamic and structural data from the literature, i.e. with the three observed s values ≈ 5, 6, 9 corresponding with the three activation energies -2.20, -2.52, -2.67 eV (cf. II-2, § 1 p.6), and with q values varying from 2 to 0 as $z$ increases. In other terms, for low values of $z$, the existence of doubly charged iron vacancies $V_{Fe}''$ is associated with a major population of holes, which corresponds to a p mode of conduction. For high values of $z$, the existence of singly charged or neutral vacancies is associated with a n mode of conduction (*i.e.*, to the injection of electrons in the conduction band). For intermediate values of $z$, an intermediate state exists with a mix of holes in the valence band and electrons in the conduction band, following a rate varying from $z_{1/2}$ to $z_{2/3}$.

These observations are supported by the p-n transition described in the literature (See (62Tan) Fig. II-10). They provide a complete interpretation for the existence of various defect configurations corresponding to pseudo-phases in the entire wüstite phase diagram. Therefore, the W1, W2 and W3 modifications can be justified from the equation (V-9) above.

For mean values of $z$, neutron diffraction data show that the clusters are assembled in limited ordered zones. For high values of $z$, the ordered cluster zones percolate (See HREM images, $z$=0.08 by (73Tch) (74Iij), $z$=0.10 (Sha))), and the change in vacancy charge as $z$ increases is accompanied by injection of free electrons in the conduction band. As temperature T increases, the size of the cluster zones decreases and the number of cluster zones increases. Correlatively, the population of free defects increases.

The W1 pseudo-phase is associated with a small number of clusters assembled in cluster zones, and a large population of free defects with numerous free $V_{Fe}''$ vacancies. In this domain, holes are the major electric charge carriers. The W3 pseudo-phase is constituted of large cluster zones covering the entire lattice adjacent to the magnetite zone. The neutralization of vacancies (or decreasing q charges) is associated with a large number of electrons occupying the conduction band. Because the pseudo-phase W2 appears to be characterized by a combination of the two thermodynamic behaviors corresponding to the pseudo-phases W1 and W3, a progressive evolution is observed at the percolation scale. There should exist an intermediate structural configuration corresponding to the formation of percolation paths between cluster zones, and the mix of conduction modes involving both holes and electrons.

## VI – Percolation approach

A primitive « composite picture » of possible evolution of the defect structure needing only two clusters and distinct populations through the subdomains was proposed by (82Gav). Nearly at the same time, a similar simple model of the defect structure was modelized in terms of the so-called theoretical Cluster Component Method (*See* Section II-2 §14 Men *et al.* and Fig. II-11).

In Figure VI-1 a schematic representation of the distribution of clusters inside cluster zones (CLZ) and cluster zones in the lattice is drawn in a likely percolation process. The size of cluster zones (CLZ) (represented by black squares) is conditioned by the equilibrium temperature. White squares represent zones with only disordered point defects (free vacancies, individual interstitials $Fe^{3+}$).

- In the W1 domain, the CLZ's are isolated and immersed in a defect lattice. The transition line W1/W2 corresponds to a specific threshold $z$ value (for fixed T) where there is a starting percolation state.

- The W2 domain might be constituted of large black zones, with percolative paths through the lattice for both types of zones. The W2/W3

transition line might be reached when the residual white zones are no more percolative.

- The W3 state is reached when black zones envelop white zones up to a limit state where the entire lattice is covered by a continuum of clusters, *i.e.*, at the boundary W3/Fe3O4.

The effect of temperature is to modify the sizes of clusters zones and the limits of composition for each boundary. As temperature increases, the cluster zones are destabilized and the size of these zones decreases.

Because of the complexity of this system, it is obvious that no unique specific model can represent the various evolutions. Nevertheless, this approach shows that the modifications W1, W2, W3 may be justified simply from changes in electrical charges of point defects, so of clusters, and are strongly correlated with composition $z$, and the degree of ordering.

**1-Low $z$ values: pseudo phase W1**

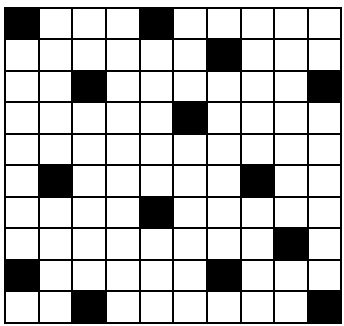

**2-Start of percolation: threshold $z$ value on the boundary W1/W2**

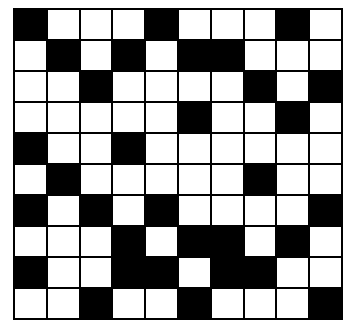

**3-Intermediate $z$ values on the boundary W2/W3 (no percolation for white zones)**

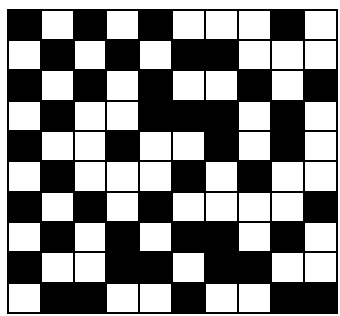

**4- Pseudo-phase W3**

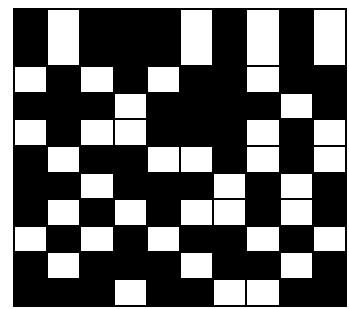

**5- Limit of existence of W3 (/Fe3O4)**

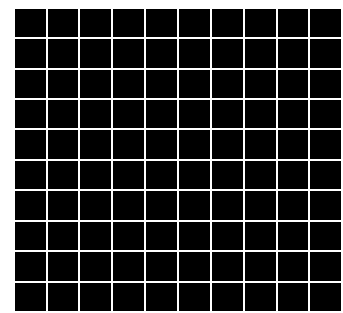

**Figure VI-1 – Schematic representation of the pseudo-phases W1, W2, W3.
A similar point of view was previously suggested in the literature (R-97Gle).**

This scheme implies nearly continuous transformations from Wi to W(i+1) at the level of the slight deviations from the main diffraction process. However, the discontinuities observed in the partial molar properties (Fig. II-6 and -7 above), and in the kinetic sequences (10Wor) at the com-positions $x_{i/j}$ for the boundaries of the subdomains correspond to abrupt and small-scale differences in internal energy and configurational entropy, due to the change of cluster type (or charge), and to the change in long-range order of clusters.

## VII - CONCLUSIONS.

**Point defects**. Up to now, the subdivision of the equilibrium phase diagram of wüstite appears to be highly complex, with the existence of several domains with different thermodynamic behaviors and high order transitions (at least of second order). The general chemical formula is based on the presence of iron vacancies noted $V_{Fe}^{q(\prime)}$ where $q^{(\prime)}$ designates the effective charge with q = 0 to 2, associated with $Fe^{3+}$ distributed on octahedral sites (fraction: *2z-t*) and tetrahedral sites (fraction: *t*) of the basic fcc lattice of NaCl type:

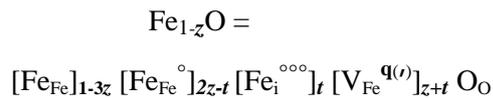

$$Fe_{1-z}O = [Fe_{Fe}]_{1-3z}\ [Fe_{Fe}^{\circ}]_{2z-t}\ [Fe_i^{\circ\circ\circ}]_t\ [V_{Fe}^{q(\prime)}]_{z+t}\ O_O$$

The electroneutrality is thought to arise from holes in the valence band and/or electrons in the conduction band.

**Clusters.** Works from several different sources clearly demonstrate the existence of clusters of defects constituted of iron vacancies coupled with cations $Fe^{3+}$ in tetrahedral interstitial sites, and mixture of $Fe^{2+}/Fe^{3+}$ cations in octahedral sites of the basic fcc lattice. These clusters can coexist with free vacancies. However, at present, the exact form of clusters noted (m:n) (m vacancies $V^{q(\prime)}$ linked to n interstitials $Fe_i^{\circ}$) is far from being determined definitively.

The ratio **R** = (*z+t*)/*t* was obtained mainly from neutron diffraction experiments in the range 2 to 4 depending on the authors. Several authors observed quasi constant R values in the range 2 to 3, while others observed a variable or constant value in the range 2.8 to 4. It should be noted that if the wüstite lattice is occupied solely by the (m:n) isolated cluster, then the ratio R should be equal to (*z+t*)/*t* = m/n.

**The pseudo-phases.** Inside the initial phase diagram, the existence of distinct areas corresponding to the pseudo phases W1, W2, W3 is supported by some experimental evidences from different authors. It is necessary to remember that the term « boundary » designates a threshold of percolation at the exact separation between two thermodynamic behaviors, corresponding to progressive evolutions of short- and long-range orderings (Fig. IV-I).

A supplementary first order transition W ⇆ W' was directly linked to the first order transition αFe ⇆ γFe at 912 °C. Up to now, the lack of systematic studies close to the isotherm at 912 °C limits the correct definition of this part of the Fe-O phase diagram, despite the availability of some advanced thermodynamic studies (81See) (86Sjö). The inter-pretation of these studies remains to be developed.

**Changes in cluster form.**

With respect to the existence of clusters (m:n) and the variation of the ratio R observed by certain authors, a first description of the pseudo-phases W1, W2, W3 was proposed in terms of structure changes of these clusters, with an increasing cluster size as *z* increased.

However, many contradictions must be noted. The historical (13:4) cluster resulting from X-ray diffraction analyses on a crystalline sample $Fe_{0.0902}O$ seems to be in contradiction with the composition of the studied wüstite and the 3X superstructure chosen to perform calculations. The ZnS-blende type (10:4) cluster could agree with neutron diffraction experiments at equilibrium and in quenched samples, and with electron microscopy images for some compositions *z* of quenched samples.

The (12:4) <110> specific cluster, proposed by Lebreton and Hobbs, could also agree with a superstructure of P″ type (5a x 5a x 10a) for a high composition $z$.

The existence of separated ordered cluster zones should agree with the quasi-constant k parameter. It should be noted that significant changes in cluster form (*e.g.*, (13:4) or (10:4) transformed into (4:1)) as $z$ or T vary were never demonstrated for the entire phase diagram. In addition, in studies of samples at equilibrium or after a quenching process, all observations of super-structure Bragg peaks or diffuse scattering due to disorder show that the k values (k close to 2.6-2.7) vary slowly with $z$, thus characterizing a quasi-constant repetition distance between clusters. If the form of clusters (m:n) changed strongly, the repetition distance k.a would also change strongly, which is not observed.

**New possible description of pseudo-phases.**

Pseudo phase W1 (low $z$ values and high T values) is characterized likely by free vacancies coexisting with a weak proportion of small zones of clusters.

Pseudo-phase W3 (high $z$ values and low T values) corresponds to clusters covering the whole lattice with a minimum distance of 2.55a between clusters and to a 5X superstructure resulting from a zigzag ordering of clusters.

Pseudo-phase W2 was identified as a solution of W1 and W3 of opposite energetic evolutions above its critical temperature of 300 °C. This pseudo-phase is constituted of zones of ordered clusters and zones of free vacancies, with a progressive build up from boundary $z_{1/2}$ to boundary $z_{2/3}$ at which the junction arises between cluster zones which could percolate.

The p to n transition might be associated with a starting of this percolation. This transition could be located in the W2 domain in a non-exclusive manner and is likely connected with a change in electrical charges of vacancies and clusters.

Finally, an additional feature can explain the observed electrical modifications. As $z$ increases, the electrical charges of vacancies ($V^{q(t)}= V″, V′$ or V) and clusters can change with q varying from q=2 (low $z$) to q=0 (high $z$), due to interactions between charges. This modification of charges q is associated with an increasing number of electrons populating the conduction band.

**Transition close to 912 °C.**

Up to now, the three varieties W'1, W'2 and W'3 associated to α-Fe below 912 °C have never been characterized directly from a structural point of view. The defect structure of these modifications analysed on the surface of the polycrystalline samples appears as being close to those for the W1, W2, and W3 pseudo-phases, as was proposed by Worral and Coley in their kinetic study, and systematic sorting of a large pattern of clusters from the literature. Nonetheless, these authors have provided new indisputable characteristics highlighting the existence of pseudo-phases within the wüstite equilibrium diagram.

———————

# Annex 1

## 1 - Two first versions of the Θ(*x*) phase diagram with three subdomains

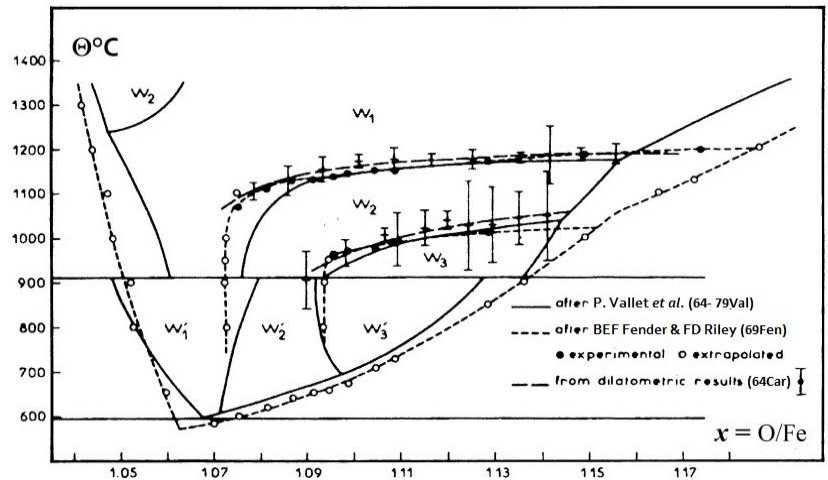

by Carel and Gavarri (76Car)

**Fig. A1-1- Comparative plot of experimental phase diagram versions by Vallet and Raccah (64Val) and Fender and Riley (69Fen).**

## 2 - Several sets from the literature of transitions located in diagram Θ[*x*]

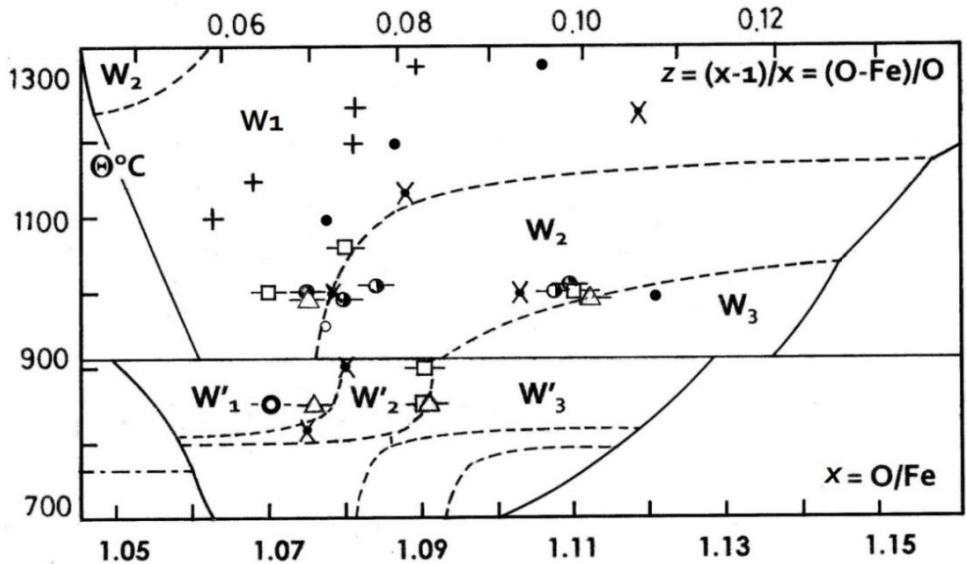

**Fig. A1-2-Transitional effects located in the diagram [Θ°C, *x*]**

a- G.H. Geiger, R.L. Levin, J.B. Wagner jr, Studies on the Defect Structure of Wüstite using Electrical Conductivity and Thermoelectric Measurements, *J. Phys. Chem. Solids* 27 (1966) 947-956, p. 955 Table 4: "*unexpected changes in the slopes of log σ vs log pO2*".

⇒ □

b- I. Bransky and A. Z. Hed, Thermogravimetric Determination of the Composition-oxygen Partial Pressure Diagram of Wustite ($Fe_{1-y}O$), *J. Amer. Ceram. Soc.- Discuss. and Notes* 51 (1968) 231-32 ; breaking points at 1000, 1100, 1200, 1300°C re-calculated for correlations of *l'* vs *z*.

⇒ ●

c- E. Riecke and K. Bohnenkamp, Über die Kinetik der Oxydation und Reduktion von Wüstit innerhalb seines Existenzgebietes, Archiv Eisenhüttenwes. N°9 (**1969**) 717-725, p.722 Fig 7; reaction rate *vs* 100*y* ($Fe_{1-y}O$) at 1000°C, pH2O/pH2, *re-analyzed* curves (I), (III) oxidation, curve (II) reduction [(I):+10°shift, (III): -10°shift], change in slope of the quasilinear segments

⇒ ◐

d- E. Takayama and N. Kimizuka (80Tak), Thermogravimetric method in the temperature range 1100°-1300°C; change in slope on the isotherms

⇒ +

e- J. Molenda, A. Stokłosa, W. Znamirowski, Transport Properties of Ferrous Oxide $Fe_{1-y}O$ at High Temperature, phys. stat. sol. (b) 142 (1987) p. 517; Fig. 2 p. 519: experimental log(σ) vs $l'$ (818-1307°C), successive segments defined from Raccah-Vallet's data to model the curvature; Fig.4: isothermal σ(y), *resulting intersections* of successive linear segments of σ($y$)

⇒ ×

f- E. J. Worral and K. S. Coley, Kinetics of the Reaction of CO2/CO Gas Mixtures with Iron Oxide, Metallurgical and Materials Transactions B 41 (**2010**) 813-823; another interpretation of Fig. 9 at 850°C gives rise to an additional point close to $x = 1.078$ as a second order type transition, identified for W'1 ↔ W'2. Another transition for $x = 1.090$ appears as a discontinuity at the location of W'2 ↔ W'3

⇒ △

**Point** ω located close to $x = 1.070$ approximate the common intersection of the three lines for re-interpreting the kinetics at 850 °C (*See* Annex 2)

⇒ ○

___________________

## Annex 2

### 1 - Separate diffusion sequences in the thermogravimetric study of the reduction in H2/H2O gas mixtures

The first experimental confirmation of the three Wi out of Vallet's team was published by (66Gei) (*See above* in Fig. A1-2). At the same time, the work of Landler and Komarek (66Lan) was available. A numerical analyze made be revealed three diffusion sequences distributed across the analysed domain (67Car). The following Fig. A2-1 was drawn as a summarized result of the determinations.

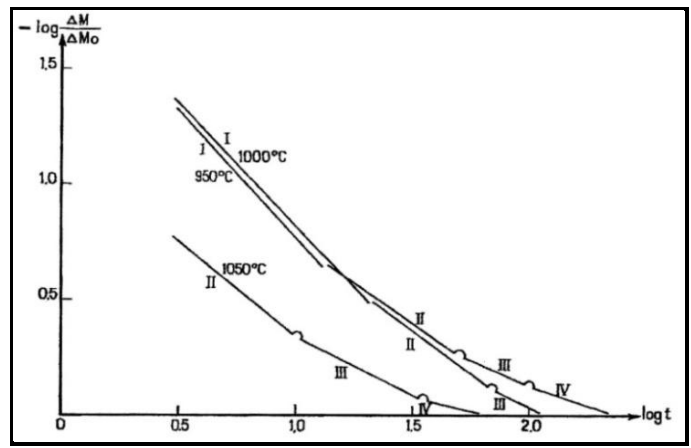

by Carel (67Car)

**Fig. A2-1 - Typical runs of reduction under H2O/H2 identified by (66Lan).**

### 2 - Re-analysis of kinetic data related to the isotherm at 850°C *by* (10Wor)

In Fig. 9 p. 821 *by* Worral and Coley (10Wor), it is possible to separate three linear variations of log $k_a$ = - m.log (CO2/CO) + log $k_0$ as displayed below in Fig. A2-2, rather than only two variations. The L. S. coefficients m = 0.48 ± 0.03 (W'1?), 0.71 ± 0.05 (W'2?), 1.03 ± 0.09 (W'3?) have been calculated. They are statistically identical to those of equations [36], [37], [38] at 1268 K by ((10Wor) p. 820). A discontinuity is observed at the location of the transition W'2 ↔ W'3 forecasted at equilibrium by (89Val). It should correspond to a sharp process (nuclear or electronic change in cluster structure, change in percolation mode, germination of a superstructure $(2.5a_o)$ x 3).

**Point** ω approximates the common graphical intersection of the three variations where the continuity for transitions W'j ↔ W'j+1 seems to be carried out. Coordinates of **point** ω are found to be log $k_a$ = -8.00558 and log $p_{CO2}/p_{CO}$ = -0.9180, *i.e.,* $l'$(ω, 1093K) = -17.38387; log p (ω, 1093K) = -12.12992. The corresponding composition would be $x = 1.069_8$ ($\cong 1.070$ at point C (86Val)). Such a convergence has been also observed $\forall$ T for $\Delta\bar{H}_O(Wi)$ (*cf* § II.2 the converging point $\Omega$ in Fig. II-5).

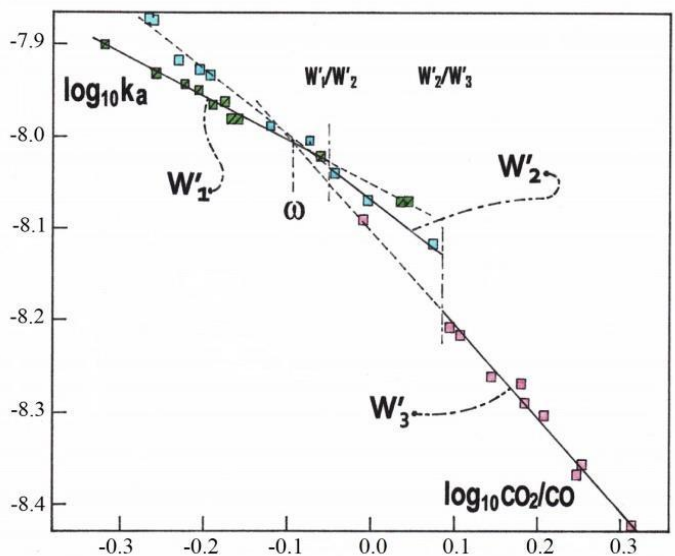

**Fig. A2-2 - Intersection W'₁/W'₂ for log CO₂/CO = -0.05052 => $x$=1.0776, expected 1.0779;
W'₂/W'₃ for log CO₂/CO= -0.08468 => $x$=1.0903, expected 1.0901;
point ω: $x$ = 1.0698 where the three variations are convergent.**

———————————————

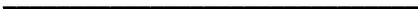